\documentclass [aps,floatfix]{revtex4}

\usepackage{amssymb}

\usepackage{epsfig}

\usepackage{graphicx}

\usepackage[centertags]{amsmath}

\usepackage{latexsym}

\usepackage{natbib}

\usepackage{picinpar}

\begin{document}

\draft

\title{Generic properties of a
quasi-one dimensional classical Wigner crystal}

\author{G. Piacente\cite{giovanni}, I.~V. Schweigert\cite{irina},
J.~J. Betouras\cite{joseph}, and F.~M. Peeters\cite{francois}}

\address{Department of Physics, University of Antwerp (UIA), B-2610 Antwerpen, Belgium}

\date{\today}

\begin{abstract}
We studied the structural, dynamical properties and melting of a
quasi-one-dimensional system of charged particles, interacting
through a screened Coulomb potential. The ground state energy was
calculated and, depending on the density and the screening length,
the system crystallizes in a number of chains. As a function of
the density (or the confining potential), the ground state
configurations and the structural transitions between them were
analyzed both by analytical and Monte Carlo calculations. The
system exhibits a rich phase diagram at zero temperature with
continuous and discontinuous structural transitions. We calculated
the normal modes of the Wigner crystal and the magneto-phonons
when an external constant magnetic field $B$ is applied. At finite
temperature the melting of the system was studied via Monte Carlo
simulations using the $modified$ $Lindemann$ $criterion$ (MLC).
The melting temperature as a function of the density was obtained
for different screening parameters. Reentrant melting as a
function of the density was found as well as evidence of
directional dependent melting. The single chain regime exhibits
anomalous melting temperatures according to the MLC and as a check
we study the pair correlation function at different densities and
different temperatures, which allowed us to formulate a different
criterion. Possible connection with recent theoretical and
experimental results are discussed and experiments are proposed.
\end{abstract}

\maketitle

\section{INTRODUCTION}

Recently there has been a great deal of interest in mesoscopic
systems consisting of interacting particles in low dimensions or
confined geometries. A class of quantum anisotropic systems
exhibiting ``stripe'' behavior appears in the quantum Hall effect
\cite{qhestripes}, in oxide manganites and high-T$_c$
superconductors \cite{hightc} where electronic strong correlations
are responsible for the formation of these inhomogeneous phases.
Another class of confined quasi one-dimensional (Q1D) geometries
appears in many diverse fields of research and some typical and
important examples from the experimental point of view are :
electrons on liquid Helium \cite{glasson,kovdrya}, microfluidic
devices \cite{whitesides}, colloidal suspensions \cite{zahn} and
confined dusty plasma \cite{chu}.

A major phenomenon which is expected to occur in charged particles
interacting via a Coulomb or screened Coulomb potential is Wigner
crystallization (WC) \cite{wigner} at low enough temperatures and
densities when the potential energy overwhelms the kinetic energy.
Indeed, evidence of such a type of transition was found very
recently \cite{glasson} in experiments on electrons on the surface
of liquid Helium where the electrons were confined by metallic
gates and exhibited dynamical ordering in the form of filaments.
This particular experiment posed many interesting questions
regarding the nature of the transition to WC, its density
dependence and the melting. Furthermore, the considered system has
been proposed as a possible step towards the realization of a
quantum computer with electrons floating on liquid Helium
\cite{platzman}.

In this paper, as a first step towards the understanding of the
behavior of these systems, we start with a two-dimensional system
consisting of an infinite number of charged particles and we
impose a parabolic confining potential in one direction. The
particles interact with a Yukawa-type potential where the
screening length is an external parameter. Physically, it can be
adjusted e.g. by the gate voltage that confines the electrons. The
combination of the interaction among particles and the external
potential leads to a rich structural phase diagram as a function
of the screening length $\lambda$ and the density $n$ of the
system. The structural units (at temperature $T = 0 K$) are
parallel chains of particles the number of which depend on the
values of $\lambda$ and $n$. The transition from one configuration
to the other can be obtained via a first or a second order
transition.

Before proceeding further, we should comment on the possibility of
two-dimensional crystalline order. According to the Mermin-Wagner
theorem \cite{Mermin} there is no true long-range crystalline
order in two dimensions. However this theorem is only strictly
valid when the potential falls off faster than $1/r$ and in the
thermodynamic limit. When the same arguments of the theorem are
applied to a large but finite system, no inconsistencies arise
from the assumption of crystalline order. Thus any system that can
be studied in laboratory or in computer simulations can exhibit
crystalline order \cite{Gann}. On the other hand short-range order
is expected to form even in the thermodynamic limit.

In a related work \cite{dubin} which discussed the temperature
equilibration of a one-dimensional Coulomb chain, two different
equilibration temperatures were assigned ($T_{\perp}$ and
$T_{||}$), reflecting the different behavior of the modes due to
the strong confinement.

The WC in strictly one dimension and in the quantum regime was
first studied by Schultz \cite{schultz}. He found that for
arbitrarily weak Coulomb interaction the density correlations at
wave vector $4k_F$ decay extremely slowly (the most slowly decay
term is $\propto\exp{(-c\sqrt{\ln{x}})}$ .

Other remarkable work on the quantum transport and pinning in the
presence of weak disorder, where it was shown that quantum
fluctuations soften the pinning barrier and charge transfer occurs
due to thermally assisted tunneling, is described in
\cite{glazman}.

In addition to the structural properties, it is instructive to
study the normal modes of these kind of anisotropic systems. There
are optical and acoustical branches and their number is equal to
the number of chains. The acoustical modes correspond to motion
along the unconfined direction and the optical ones to motion
along the confined direction. There is softening of an optical
phonon at those values of the density for which we have a
continuous structural transition. We also study the collective
excitations in the presence of a constant magnetic field
perpendicular to the plane of the system. These modes
(magnetophonons), can be directly detected experimentally
\cite{magn1,magn2}.

Another important aspect of the problem is the melting as the
temperature is raised. The mechanisms of melting is of great
scientific and technological importance. In infinite
two-dimensional (2D) systems theory \cite{kthny}, based on
unbinding of defects, predicts a two-stage melting where the two
stages are continuous. Recent theoretical studies of melting of
colloidal crystals in the presence of a one-dimensional periodic
potential \cite{radzihovsky} revealed a number of novel phases and
the possibility of reentrant melting. These results depend on the
commensurability ratio $p=a/d_{ext}$ where $a$ is the spacing
between the Bragg planes of the 2D system and $d_{ext}$ the period
of the external periodic potential. This kind of system was
realized experimentally \cite{wei} in 2D colloids in the presence
of two interfering laser beams. The present work is complementary
to the work of Frey, Nelson and Radzihovsky \cite{radzihovsky} in
the sense that a single confining potential is considered here,
which is not repeated in space. Therefore it can be viewed as a
study of a focused portion of the infinite 2D system, where we pay
attention to only one potential trough neglecting the interaction
with the rest. With respect to the melting, we found the following
remarkable results:(i) a phase diagram which exhibits reentrant
melting behavior as a function of the density where the different
configurations are explored, (ii) a regime of frustration exists
close to the structural transitions, and (iii) there is evidence
that the system first melts in the unconfined direction and
subsequently in the direction where it is confined exhibiting a
regime similar to the {\it locked floating solid} regime found in
Ref. \cite{radzihovsky}.

The paper is organized as follows. In Sec. II, we present the
model and the methods used. In Sec. III we study the zero
temperature phase diagram and properties of the structural
transitions. Sec. IV is devoted to the study of the normal modes
of the system and in the presence of, or without, an external
magnetic field $B$. In Sec. V we study the melting and analyze
furthermore in some details the problem of the single chain
melting.

Finally we discuss the connections with recent experimental
results and suggest experiments where this behavior can be
observed in Sec. VI.

\section{MODEL AND  METHODS}

The system is modelled by an infinite number of classical charged
particles with identical charge $q$, moving in a plane with
coordinates $\vec{r}=(x,y)$. The particles interact through a
Yukawa potential and an additional parabolic potential confines
the particle motion in the $y$ direction. The Hamiltonian of the
system is given by :
\begin{equation}
H=\frac{q^2}{\epsilon} \sum_{i \neq j}
\frac{exp(-|\vec{r_i}-\vec{r_j}|/\lambda)}{|\vec{r_i}-\vec{r_j}|}
+ \sum_i \frac{1}{2} m {\omega_0}^2 {y_i}^2
\end{equation}
\noindent where $m$ is the mass of each particle, $\epsilon$ is
the dielectric constant of the medium particles are moving in,
$\omega_0$ measures the strength of the confining potential. The
Hamiltonian can be rewritten in a dimensionless form, introducing
the quantities $r_{0}=(2q/m\varepsilon \omega_{0}^{2})^{1/3}$ as
unit of length and $E_{0}=(m\omega _{0}^{2}q^{4}/2\varepsilon
^{2})^{1/3}$ as unit of energy. Then it takes the form
\begin{equation}
H'=\sum_{i \neq j} \frac{exp(-\kappa |\vec{r'_i}-\vec{r'_j}|)}
{|\vec{r'_i}-\vec{r'_j}|} + \sum_i {y'_i}^2
\end{equation}
\noindent where $H^{\prime }=H/E_{0}$, $\kappa =r_{0}/\lambda $
and $\overrightarrow{r}^{\prime }=\overrightarrow{r}/r_{0}$. This
transformation is particularly interesting because now the
Hamiltonian no longer depends on the specifics of the system and
becomes only a function of the density and the dimensionless
inverse screening length. The quantities introduced allow us to
define a dimensionless temperature $T^{\prime }=T/T_{0}$ with
$T_{0}=(m\omega _{0}^{2}q^{4}/2\varepsilon^{2})^{1/3}k_{B.}^{-1}$.

For the calculations of the ground state energy we used a
combination of analytical calculations and Monte Carlo simulations
with the standard Metropolis algorithm. This recursive algorithm
consists in displacing randomly one particle and accepting the new
configuration if its energy is lower than the previous one; if the
new configuration has a larger energy the displacements are
accepted with probability $\delta <\exp (-\Delta E/T)$, where $
\delta $ is a random number between $0$  and $1$ and $\Delta E$ is
the increment in the energy. We have allowed the system to
approach its equilibrium state at some temperature $T$, after
executing $10^{5}\div 10^{6}$ Monte Carlo steps. We have used the
technique of simulated annealing to reach the $T=0$ equilibrium
configuration: first the system has been heated up and then cooled
down to a very low temperature. In the simulations typically 300
particles were used and in order to simulate an infinitely long
system periodical boundary conditions (Born-Von Karman) were
introduced.

\section{Ground state configurations}

\subsection{Phase Diagram}

The charged particles crystallize in a certain number of chains.
Each chain has the same density resulting in a total
one-dimensional density $\widetilde{n}_{e}$. It is then possible
to calculate the energy per particle for each configuration and to
check the favored one as a function of the parameters of the
system. If $a$ is the separation between two adjacent particles in
the same chain, we can define the dimensionless  linear density $
\widetilde{n}_{e}=lr_{0}/a$, where $l$ is the number of chains.




In the case of multiple chains, in order to have a better packing
(or in other words to minimize the interaction energy by
maximizing the separation among particles in different chains),
the chains are staggered with respect to each other by $a/2$ in
the $x$ direction. In an infinite lattice this will lead to the
hexagonal WC \cite{bonsall}. We calculated the energy per particle
as a function of the density for the first six possible
configurations of the system.

If the particles crystallize in a single chain, the minimum energy
is obtained when the particles are placed on the $y$ axis, where
the confining potential is zero. In this case the linear density
is $\widetilde{n}_{e}=r_{0}/a$ and the $x$ coordinate of the
particles are $x_{i}=ia$, with $i=0,\pm 1,\pm 2,...,\pm \infty $.
The energy per particle is :

\begin{equation}
E_{1}=\widetilde{n}_{e}\underset{j=1}{\overset{\infty }{\sum
}}\frac{1}{j} \exp(-\kappa j/\widetilde{n}_{e}).
\end{equation}
\noindent The case of Coulomb interaction is treated using the
Ewald summation method so that the summation over long distance
can be done effectively. Following the standard procedure
\cite{bonsall} we obtain for $E_1$ :
\begin{equation}
E_1 (\kappa=0) = \frac{{\widetilde{n}_{e}}^2}{ \sqrt{\pi}}
{\underset{x \rightarrow 0}{lim}} [\underset{j}\sum 2 e^{-2 \pi j
x} {\Phi}_1(j \pi /2{\widetilde{n}_e}) + \underset{j \neq 0} \sum
{\Phi}_2({\widetilde{n}_e}^2 (x-j)^2) +
\frac{1}{\widetilde{n}_{e}}{\Phi}_2({\widetilde{n}_e}^2 x^2)
-\frac{\sqrt{\pi}}{{\widetilde{n}_e}^2} \frac{1}{x}],
\end{equation}
\noindent where ${\Phi}_1(x) = \sqrt{\pi} \int_{x}^{\infty} dt \;
exp(-t^2) \; 1/t$, ${\Phi}_2 (x) = \sqrt{\frac{\pi}{x}}$
erfc$(\sqrt{x})$ and erfc$(y)=1- \frac{2}{\sqrt{\pi}}\int_0^y
e^{-t^2} dt$.

The first summation contains a divergent term at $j=0$ coming from
the lower limit of the integration in the function
${\Phi}_1(x=0)$. This divergence is remedied if we subtract the
interaction energy $(E_b)$ of the negatively charged particles
with the positive background which also diverges logarithmically
in one dimension. In that case we can proceed using the limit
${\underset{x \rightarrow 0}{lim}} x^{-1}$ erf$(x) =
2/\sqrt{\pi}$:
\begin{equation}
\Delta E_1 = E_1(\kappa=0) - E_b =
\frac{{\widetilde{n}_{e}}^2}{\sqrt{\pi}} [\underset{j \neq 0} \sum
2{\Phi}_1(j \pi /2{\widetilde{n}_e}) + \underset{j \neq 0} \sum
{\Phi}_2({\widetilde{n}_e}^2 j^2)] - \frac{2}{\sqrt{\pi}}
\widetilde{n}_e.
\end{equation}
In the two-chain configuration the particles crystallize in two
parallel lines separated by a distance $d$ and displaced by a
distance $a/2$ along the $y$-axis. The energy per particle in this
case is:
\begin{equation}
E_{2}=\frac{c^{2}}{\widetilde{n}_{e}^{2}}
+\frac{\widetilde{n}_{e}}{2}\underset{j=1}{\overset{\infty }{\sum
}}\frac{1}{j} \exp (-2\kappa
j/\widetilde{n}_{e})+\frac{\widetilde{n}_{e}}{2}\underset{j=1}{\overset{\infty
}{\sum }}\frac{\exp(-2\kappa
\sqrt{(j-1/2)^{2}+c^{2}}/\widetilde{n}_{e})}{\sqrt{(j-1/2)^{2}+c^{2}}},
\end{equation}
\noindent where $\widetilde{n}_{e}=2r_{0}/a$ and $c=d/a$. The
first term in (4) is the potential energy due to the confining
potential, the second term is the energy due to the intra-chain
interaction and the last term represents the inter-chain
interactions. Minimizing $E_{2}$ with respect to the separation
between the chains, $c$, we obtained the ground state energy for
the two-chain configuration.

Similar straightforward but tedious calculations were done for the
other multi-chain structures. By symmetry there is one intrachain
distance in the three-chain structure, two in the four- and
five-chain structures and three in the six-chain structure. The
corresponding expressions for the energy are relegated, for
completeness, to Appendix A.

Calculating the energy minimum for each configuration for
different values of $\widetilde{n}_{e}$  at fixed  $\kappa $, we
obtain the energy per particle $E$. In Fig. 1 we show $E$ as a
function of the density $\widetilde{n}_e$ for $\kappa =1$. Notice
that for certain density ranges more than one configuration can be
stable (this is made more clear in the insets of Fig. 1 for
$\widetilde{n}_e$ around 2 and 4.7). In the low density limit the
energy per particle is given by the first term of Eq. (3) :
$E=\widetilde{n}_{e}  exp(-\kappa /\widetilde{n}_e$), while the
rest of the curve can be fitted to $E=-0.0194{\widetilde{n}_e}^{2}
+ 0.720 \widetilde{n}_e - 0.245$ with an error less than 2.3\%.

Calculating the energy minima  for different  $\widetilde{n}_{e}$
and different $\kappa$ we obtain the zero temperature phase
diagram of Fig. 2. For $\kappa=0$ we recover the Coulomb limit. We
found that the energy  obtained by the analytical method is in
excellent agreement with the one obtained by our Monte Carlo
simulations with a difference between them less than 0.3 \%.

We observe the following sequence of transitions as the density
increases : from one to the two-chain structure then to the
four-chain configuration,  back to the three-chain and again to
four and then to five, six etc. Notice the remarkable fact that
between the 2 and 3 chain configuration there is a small
intermediate region where a four-chain configuration has a lower
energy. For all other transitions the number of chains increases
only by one unit, i.e. $n \rightarrow n+1$. The relative lateral
position of the different chains are depicted in Fig. 3 as a
function of the density $\widetilde{n}_e$. In the case of two and
three-chain the inter-chain distance increases as the density
increases. This is also true for the four-chain configuration too,
with some differences. In the first four-chain regime of the phase
diagram, the distance between the two internal chains is larger
than the distance between the internal chains and the external
ones, in the second regime the behavior of the system is the
opposite with the distance between internal and external chains
larger than the one between internal chains. For the other
structures the interchain distance is always a growing function of
the density. It is evident that only the first transition is
continuous with a clear bifurcation.

In order to gain some insight on the distribution of the energy in
this anisotropic system we present in Fig. 4 the energy per
particle for each chain. This is computed by considering a
particle at a particular chain and taking into account all the
interactions with the rest of the particles. The cases of interest
are the configurations for which it is possible to distinguish
internal from external chains and may be related to the difference
in the melting behavior which is discussed in Sec. V. The
interesting observation is that in every case the energy per
particle is larger in the external chain than the internal ones.

This asymmetry reflects the fact that for each particle residing
in an external chain the gain in energy due to the confining
potential is higher than the difference in the Coulomb energy due
to the lack of symmetric neighboring chains, as compared to a
particle residing in an internal chain. E.g. for a three-chain
system where the middle chain is the 0th and the external ones are
denoted by +1 and -1,  we have for the energy of two particles :
\begin{equation}
E_{\pm 1} - E_{0} = E_{conf,\pm 1}+ E_{Coulomb,+1,-1}-
E_{Coulomb,\pm 1,0} > 0,
\end{equation}
\noindent where E$_{Coulomb,\alpha,\beta}$ denotes the Coulomb
energy of a particle residing in chain $\alpha$ interacting with
the particles in chain $\beta$ and E$_{conf,\alpha}$ denotes its
confining energy.

In the case of the first density regimes where the four-chain
structure is optimal this difference is not large due to the fact
that the internal distance is less than the external. On the
contrary, the difference is much larger in the second regime of
the four-chain structure. Another interesting observation is that
as we approach the limit of Coulomb interactions ($\kappa \ll 1$)
the energy difference tends to vanish and the system behaves
isotropically.

\subsection{Structural transitions}

We have seen that by increasing the density, the system changes
its configuration, in other words it undergoes a ``structural
transition''. It is a natural question to study the order of these
transitions. For this purpose the derivative of the energy with
respect to the density was calculated which is shown in Fig. 5 for
the case of $\kappa = 1$. Only the transition between the one and
the two-chain configuration is continuous and all the others are
discontinuous. This conclusion agrees with the results of Fig. 3
where discontinuous changes of the lateral position of the
particles correspond to first order transitions. The transition $1
\rightarrow 2$ is a \textquotedblleft zig-zag\textquotedblright
transition \cite{zigzag} (Fig. 6). The transition $2 \rightarrow
4$ occurs through a \textquotedblleft zig-zag\textquotedblright
transition of each of the two chains accompanied by a shift of
$a/4$ along the chain, which makes it a weakly discontinuous
transition (Fig. 6). In principle, these kind of almost
``zig-zag'' transitions are possible for three-, four-, five- and
six- chains to result into six-, eight-, ten- and twelve- chain
structures, respectively. Actually, these were observed during the
numerical simulations, especially for very small value of
$\kappa$, but they represent metastable states and are not the
most energetically favored configurations.

\subsection{Limit of short range interaction and large density}

In order to make the connection with the regime where the hard
core potential can be used as a working hypothesis, we investigate
the limit $\lambda \ll a$. It can be shown that the variation of
the distances between chains can be neglected and in the limit
where $m{\omega_0}^2 W^2 \ll q^2/(a \epsilon)$ ($W$ is the width
of the strip), following the spirit of the hydrodynamic
consideration of Koulakov and Shklovskii \cite{koulakov} the
difference in the distance between chains at the borders $d(\pm
W/2)$ and at the center $d(0)$ follows the relation:
\begin{equation}
\delta d_0 = d_0(W/2) - d_0(0) \approx \lambda lnl,
\end{equation}
\noindent where $l$ is the number of chains and $d_0=\sqrt{d^2+a^2/4}$.

This can be estimated by considering the pressure $\sigma_{yy}$ in
the crystal exercised by the external potential. Adopting a method
similar to Ref.\cite{koulakov,landau}:
\begin{equation}
\sigma_{yy}= -S(\sigma)\frac{m{\omega_0}^2}{2}(\frac{W^2}{4}-y^2),
\end{equation}
\noindent where $3/4 \leq S(\sigma) \leq  1$ and $\sigma$ is the
Poisson ratio. We assume a uniform density $n$ and $S(\sigma)
\approx 1$. Then, balancing the force by the pressure and the
interaction forces we get (in this estimate we keep the dimensions
for clarity):
\begin{equation}
\frac{2dq^2}{\epsilon(d^2+a^2/4)}exp(-\sqrt{d^2+a^2/4}/\lambda)\sim
\frac{n d m{\omega_0}^2}{2} (W^2/4-y^2) ,
\end{equation}
\noindent from this relation :
\begin{equation}
d_0(y) \equiv \sqrt{d(y)^2+a^2/4}\approx \lambda
ln[\frac{4q^2}{\epsilon m {\omega_0}^2 n
(W^2/4-y^2)(d(y)^2+a^2/4)}] ,
\end{equation}
\noindent subtracting the values of $d_0$ at $W-a/2$ and 0 we
obtain Eq. (6).

Therefore in the case of very short-range interaction $\delta d
\ll d(0)$. Then one can adopt the hard core potential and
essentially the total energy becomes the sum of the energy of each
particle due to the confining potential. The average energy per
unit length E/L then reads :
\begin{equation}
E/L \approx \frac{1}{24} m {\omega_0}^2 \frac{W^2}{l a} .
\end{equation}

\section{NORMAL MODES}

\subsection{Normal modes in the absence of an external
magnetic field}

We now turn to the calculation of the normal modes of the system,
following the standard harmonic approximation \cite{pines} and
exploiting the translational invariance of the system along the
$x$ direction. The number of chains determines the number of
particles in each unit cell and therefore the number of degrees of
freedom per unit cell. So if $l$ is the number of chains there
will be $2l$ branches for the normal mode dispersion curves: $l$
acoustical branches as well as $l$ optical ones.  Note that for
ordinary bidimensional crystals there are 2 acoustical branches
and $2r - 2$ optical branches, if $r$ is the number of atomic
species in the unit cell. We present the results for the one, two
and three-chain structure in Fig. 7. Note that for the one chain
structure the unit cell consists of a single particle, i.e. $r=1$,
and therefore one expects only a single acoustical branch and no
optical branch. The appearance of the optical branch is a
consequence of the presence of the confining potential in the
$y$-direction. Note that for $k \rightarrow 0$, $\omega_{opt}
\approx \omega_0$, which corresponds to the center of mass motion
of the system in the confining potential.

In order to find the eigenmodes we solve the system of equations :
\begin{equation}
(\omega^{2}\delta_{\alpha\beta,ij}-D_{\alpha\beta,ij})Q_{\beta,j}=0,
\end{equation}
\noindent where $Q_{\beta,j}$ is the displacement of the particle
$j$ from its equilibrium position in the $\beta$ direction,
$(\alpha,\beta)\equiv{(x,y)}$, $\delta_{\alpha\beta,ij}$ is a unit
matrix and $D_{\alpha\beta,ij}$ is the dynamical matrix defined
by:
\begin{equation}
D_{\alpha\beta,ij} = \frac{1}{m} \sum_{\nu}
\phi_{\alpha,\beta}(\nu) e^{-i \nu qa}
\end{equation}

\noindent where $\nu$ is an integer assigned to each unit cell and
the force constants are :
\begin{equation}
\phi_{\alpha,\beta}(\nu) = \partial_{\alpha} \partial_{\beta}
\frac{exp(-\kappa\sqrt{(x-x')^2+(y-y')^2}}{\sqrt{(x-x')^2+(y-y')^2}},
\;\;\nu \neq 0,
\end{equation}
\noindent evaluated at $x-x' \in  \{a \nu, a (\nu+1/2)\}$, $y-y'=$
relevant interchain distance. And
\begin{equation}
\phi_{\alpha,\beta}(\nu=0) = - \sum_{\nu \neq 0}
\phi_{\alpha,\beta}(\nu).
\end{equation}
\noindent All the frequencies are measured in unit of
$\omega_{0}/\sqrt{2}$. In Appendix B we present for completeness
the expressions for the matrix where as an example the modes for
the three-chain structure were calculated.

The main feature is the softening of the optical mode of the one
chain structure at the values of $\widetilde{n}_{e}$ and $\kappa$
where the structural transition is observed (``zig-zag''
transition) accompanied by a hardening of the acoustical branch
(Fig. 8), which confirms that $1\rightarrow 2$ is a continuous
transition as asserted before.

Studying the eigenvectors of the dynamical matrix it is easy to
recognize that the optical modes are identified with the motion in
the direction of confinement ($y$ direction) while the acoustical
modes are identified with the motion in the unconfined $x$
direction.

The eigenfrequencies for the single chain are given by :
$\omega_{ac} = \sqrt{A_1}$ for the acoustical branch and
$\omega_{opt} = \sqrt{1+A_2}$ for the optical branch, where $A_1$
and $A_2$ are defined in Appendix B.

In the limit of small wavenumbers $k$, the summations can be done
analytically and we obtain:
\begin{eqnarray}
\omega_{ac}(k) &=& [ - ln(1-e^{-\kappa/\widetilde{n}_e}) +
\frac{\kappa} {\widetilde{n}_e} \frac{e^{-\kappa/\widetilde{n}_e}}
{1-e^{-\kappa/\widetilde{n}_e}} + \frac{{\kappa}^2}{2
{\widetilde{n}_e}^2}\frac{e^{-\kappa/\widetilde{n}_e}}
{(1-e^{-\kappa/\widetilde{n}_e})^2}]^{1/2} {\widetilde{n}_e}^{3/2} |k|a, \\
\omega_{opt}(k) &=& \{1 - [-ln(1-e^{-\kappa/\widetilde{n}_e}) +
\frac{\kappa} {\widetilde{n}_e} \frac{e^{-\kappa/\widetilde{n}_e}}
{1-e^{-\kappa/\widetilde{n}_e}}] {\widetilde{n}_e}^3 k^2 a^2
\}^{1/2},
\end{eqnarray}
\noindent which gives explicitly the dependence of the modes on
the density and the screening parameter. In the limit
$\kappa/\widetilde{n}_e \gg 1$ :

\begin{eqnarray}
\omega_{ac}(k) &=& e^{-\kappa/\widetilde{n}_e} \frac{{\kappa}^2}
{2\sqrt{\widetilde{n}_e}} |k|a , \\
\omega_{opt}(k) &=& 1-  e^{-\kappa/\widetilde{n}_e}
\frac{{\widetilde{n}_e}^2}{2\kappa} k^2 a^2,
\end{eqnarray}

\noindent while in the opposite limit $\kappa/\widetilde{n}_e
\ll1$ :
\begin{eqnarray}
\omega_{ac}(k) &=& [\frac{3}{2} +
ln(\frac{\widetilde{n}_e}{\kappa})]^{1/2}
{\widetilde{n}_e}^{3/2} |k|a , \\
\omega_{opt}(k) &=& \{ 1 -[ 1+ln(\frac{\widetilde{n}_e}{\kappa})]
{\widetilde{n}_e}^3 k^2 a^2 \}^{1/2}.
\end{eqnarray}

There is a remarkable difference in the optical branch of the
spectrum between the single chain and the two and three-chain
structures. In the first case the frequency of the optical branch
decreases as the wavenumber $k$ increases, while for the two and
three-chain structures the optical frequency increases. In the
single chain configuration the optical mode corresponds to
oscillations of the particles in the confined direction (see e.g.
Fig. 9(b)) which reduces the Coulomb repulsive energy. For the two
chain configuration the normal modes are shown in Fig. 9(c-g). In
fact this branch is nothing else than a transverse acoustical mode
while the acoustical branch corresponds to longitudinal motion
\cite{kovdrya,sokolov}.

\subsection{Normal modes in the presence of an external
magnetic field}

We now consider the effect of applying a constant magnetic field
$B$ in the $z$ direction. For quantum particles, the magnetic
field can localize the charged particles into cyclotron orbits,
therefore aiding the formation of a Wigner crystal in the presence
of a magnetic field. It is known \cite{vanlee} that in a classical
system an external magnetic field does not alter the statistical
properties of the system and consequently the structural
properties and the melting temperature are insensitive to the
magnetic field strength. But on the other hand the character of
motion of the particles is altered significantly when the
cyclotron frequency is larger than the eigenfrequencies of the
system. The magneto-phonon spectrum of an infinite 2D Wigner
crystal in a magnetic field was obtained in \cite{chaplik},
\cite{bonsall}. In the presence of $B$, the system of equations is
modified to:
\begin{equation}
(\omega^{2}\delta_{\alpha\beta,ij}-D_{\alpha\beta,ij}+i\omega\omega_{c}
\xi_{\alpha\beta}\delta_{ij})Q_{\beta,j}=0
\end{equation}
\noindent where $\xi_{\alpha\beta}$ is the Levi-Civita tensor and
$\omega_{c} = qB/mc$ is the cyclotron frequency. In Fig. 10 we
show some typical dispersion curves for the one and three-chain
structures for different values of $\omega_{c}$. It is interesting
to notice how the optical modes couple with the magnetic field,
the optical frequencies follow the cyclotron frequency and for
very high field strength there is no significant difference
between $\omega_{opt}$ and $\omega_{c}$. The acoustical
frequencies on the other hand decrease with the magnetic field
strength. For the single chain the eigenfrequencies are modified
to:

\begin{equation}
\omega(k) = \{\frac{1}{2} (1 + A_1 + A_2 + {\omega_c}^2) \pm
\frac{1}{2} [(1 + A_1 + A_2 + {\omega_c}^2)^2 - 4
A_1(1+A_2)]^{1/2}\}^{1/2}
\end{equation}

\noindent where $A_1$ and $A_2$ are given in Appendix B. For very
large field when $\omega_c \gg$ \{$A_1$,$A_2$,1\} the gap between
the optical branches and the acoustical ones approaches
$\omega_{c}$. The optical frequency reflects the cyclotron motion
of the system which supresses any soft excitation.
 As before, it
is interesting to study the normal modes at the critical density
of the transition from the one-chain to the two-chain structure
(Fig. 11). We observe that there is always softening at the same
density, independently of the strength of the magnetic field, but
with a main difference that for zero magnetic field strength the
modes which soften is the optical one while when the magnetic
field strength is nonzero, the acoustic mode is the one that
softens. The magnetic field induces a coupling between the
acoustical and the optical modes and there is an anticrossing
between the two branches.
 Although these findings confirm the
previous assertion that the presence of $B$ does not alter the
structural properties of the system it also reveals the
differences (softening of the acoustic mode at the same density,
influence on the gap between optical and acoustical branches and
on eigenfrequencies within each branch) which are induced by the
magnetic field.

\section{MELTING}

\subsection{General discussion and results}

In this section we study the melting of the WC by MC simulations.
After the ground state configuration was achieved as explained in
Sec. II, the system was heated up by steps of size $\Delta T$,
typically $\Delta T=5\times10^{-4}$, and equilibrated to this new
temperature during $10^{5}\div 10^{6}$ MC steps. In Fig. 12 we
show typical trajectories of particles as they arise from our MC
simulation. It is evident that there is a different behavior of
the system in the $x$ and the $y$ directions as may be expected by
the anisotropy in the two directions. In order to quantify the
observations, we studied first the potential energy as a function
of temperature (Fig. 13). In the crystalline state the potential
energy of the system increases practically linearly with
temperature and then exhibits a very fast increase in a small
critical temperature range after which it starts to increase
linearly again but now with a slightly larger slope. In the latter
region the system is in the disordered ($i. e.$ liquid) phase. The
fast increase of the potential energy is indicative of the melting
of the WC.
 To
find the critical temperatures we studied, following the spirit of
Ref. \cite{bedanov}, the modified Lindemann parameter $L_p=\langle
u^{2} \rangle/ {d_r}^{2}$, where $\langle u^{2} \rangle$ is
defined by the difference in the mean square displacements of
neighboring particles from their equilibrium sites $\vec{r}_0$ and
$d_r$ is the relevant interparticle distance as we discuss below.
The quantity $\langle u^{2} \rangle$ can be written as:

\begin{equation}
\langle u^{2}\rangle =
\frac{1}{N}\langle\underset{i=1}{\overset{N}{\sum
}}\frac{1}{N_{nb}}\underset{j=1}{\overset{N_{nb}}{\sum
}}[(\vec{r}_i-\vec{r}_{0i})-(\vec{r}_j-\vec{r}_{0j})]^{2}\rangle
\end{equation}

\noindent where $ \langle $  $\rangle $ means the average over the
MC steps, $N$ is the total number of particles in our simulation
unit cell and the index $j$ denotes the $N_{nb}$ nearest neighbors
of particle $i$. In order to describe more accurately the
difference between the two directions, we studied separately
$\langle u_{x}^{2} \rangle$ and $\langle u_{y}^{2} \rangle$ as
function of temperature. For the melting along the $x$ direction,
the distance $d_r$ is the interparticle distance $a$ introduced in
Sec. I while for melting along the $y$ direction $d_r$ is the
interchain distance which is a function of the density
$\widetilde{n}_{e}$.

At low temperatures, the mean square relative displacements slowly
increases linearly with temperature as a consequence of harmonic
oscillations of the particles about their equilibrium positions
(see Fig. 14). From Fig. 14 we notice clearly that this linear
increase is larger in the unconfined direction than in the
confined direction. In some critical temperature region, $\langle
u_{x}^{2} \rangle$ and $\langle u_{y}^{2} \rangle$ start to
increase very rapidly which is the consequence of the fact that
the particles have attained sufficient thermal energy that they
can jump between different crystallographic positions. According
to the modified Lindemann criterion MLC, when $L_p$ reaches the
(semi-empirical) critical value 0.1 the system melts. This
criterion was used to define the melting temperature $T_m$.

From the corresponding analysis two different melting
temperatures, $T_x$ and $T_y$, can be assigned. The results are
summarized in the phase diagram of Figs. 15(a-c) for
$\kappa=$0.01, 1 and 3, respectively. There are several
interesting features in these phase diagrams: ($a$) the nearly
Coulomb system ($\kappa=0.01$) has a melting temperature which is
on average 15-20 \% higher than for the screened Coulomb
inter-particle interaction with $\kappa = 1$, which has on its
turn an average melting temperature about 15 \% higher than the
screened Coulomb system with $\kappa=3$. Therefore, we conclude
that the effect of screeening is to reduce the melting
temperatures ; ($b$) a {\it reentrant} behavior is observed as a
function of density, the minima of the melting temperatures occur
at the values of the density where the structural phase
transitions were predicted (see Fig. 2); ($c$) there is a regime
close to each structural transition point where the system is {\it
frustrated}, in the sense that it fluctuates between the two
structures. In this regime, which we term as {\it frustration
regime}, the system makes continuous transitions from one
metastable state to the other which strongly reduces the melting
temperature; ($d$) for $\kappa = 1$ and $\kappa=3$, there is a
region in density for which the system melts first in the
unconfined direction while it is not melted in the confined one.
This regime resembles the findings of Ref. \cite{radzihovsky} in
the regime termed as {\it locked floating solid}. For the Coulomb
limit there is no evidence of anisotropic melting within the error
bars of our simulation. The system behaves more isotropic; ($e$)
the first four-chain regime (see insets of Figs. 15a-c) is
unstable with respect to temperature fluctuations as it is
reflected in the relative low melting temperature. In this region,
melting occurs first in the confined direction as a consequence of
the particular structural properties -the distance between the two
internal chains is larger than the distance between an internal
chain and the adjacent external one- which makes the system
unstable in the $y$ direction. In the rest of the diagram there is
evidence that the melting either starts from the unconfined
direction (e.g. it is clear in the single chain and in the low
density limit of the two-chains) or the system melts
simultaneously in both directions. ($f$) the single-chain
structure shows a relatively large melting temperature as obtained
by the MLC and deserves more attention. The study of the
single-chain is therefore postponed to the next subsection.

Furthermore, note that the MLC only takes into account the
displacement of the particles relative to the position of their
neighbors and consequently is only a measure of the local order of
the system.

Another natural question that arises is whether there is
anisotropic melting with respect to external and internal chains
in the multi-chain structures or in other words if melting starts
from the edges as observed in the experiment of Ref.
\cite{glasson} with electrons on liquid helium. The number of
filaments that were observed in the experiment was approximately
20; we have simulated the trajectories of some multi-chain
structures and the results are presented in Fig. 18. In this
picture it is clear that the most external chains are already
melted while the internal ones are still ordered. Edge melting has
also been demonstrated in the presence of a strong magnetic field
in Ref. \cite{cote} using Hartree-Fock calculations in a
two-dimensional Wigner crystal with edges. With the aid of many
numerical simulations of multi-chain systems at different
densities we observed that this kind of melting is present in our
system when the density is close enough to the critical density of
a structural transition. Close to the structural transition many
metastable states appear with a different number of particles per
chain, that is in the most external chains there are less
particles than in the internal ones. Thus the particles at the
most external chains have larger displacements from their
equilibrium positions in order to attain the stability of the
structure. Furthermore, we calculated the average root mean square
displacements of the particles from their equilibrium position
chain by chain and also $\langle u_{x}^{2} \rangle$ chain by chain
and we actually noted that these quantities are slightly larger
for external chains at temperatures

below the critical one. In Fig. 19 we present the temperature
dependence of the standard deviation ${s_x}^2=\langle
(u_{x}-\langle u_{x}\rangle)^{2}\rangle$ and ${s_x}^2=\langle
(u_{y}-\langle u_{y}\rangle)^{2}\rangle$ for the external and
internal chains in the four-chain structure. It is evident that
the position of the particles at the edges fluctuates
substantially more than the particles at the interior. We can
conjecture that, according to this physical picture, melting can
start from the edges. However, for up to the six-chain
configuration for each chain the quantities $\langle u_{x}^{2}
\rangle$ reached the critical value, approximately, all at the
same temperature. Probably, going to a larger number of filaments
one can well appreciate a different melting temperature for
external and internal chains. Finally, the chain configuration as
well as the melting which starts from the direction of the chains
is supported also by molecular dynamics simulations of the flow of
electrons in Q1D channels \cite{mehrotra}.

\subsection{Melting of the single-chain}

In Figs. 15(a-c) we observe a rather high melting temperature in
case of the one-chain structure. The origin of this behavior can
be traced back to the fact that the MLC takes into account a
larger contribution from jumps of particles between
crystallographic positions which for the single chain structure
occurs only at extremely high temperature. For the single-chain
case the jumps can only occur along the chain which requires a
larger energy than jumps of particles between different chains.

To have a better insight we investigated the behavior of $L_p$ for
different densities (Fig. 18). We notice that in the low density
limit (see Fig. 18(a)), $L_p$ $\approx$ 0.1 is reached in a region
in which there is only a gradual increase in $\langle u_{x}^{2}
\rangle$ which is very different from the multi-chain case (see
Fig. 14). Furthermore, $\langle u_{x}^{2} \rangle$ exhibits a
sublinear temperature increase.

This calls for the use of other possible criteria in order to
clarify the situation. On the other hand if the density is
relatively high (see Fig. 18), a fast increase is observed
signaling a clear melting of the system. The transition from a low
temperature linear to sublinear behaviour occurs for
$\widetilde{n}_e \approx 0.4$.

To shine light into the posed questions we studied also the pair
correlation function at different densities and temperatures, as
defined by
\begin{equation}
g(x) = \frac{L}{N^2} \sum_{i \neq j} <\delta[x-(x_i-x_j)]>,
\end{equation}
\noindent where in the summation over $N$ particles in a system of
length $L$, the diagonal terms ($i=j$) are excluded . The results
are reported in Fig. 19. It is rather evident that the melting
temperature is substantially smaller than the one obtained from
the MLC. In order to better quantify the melting temperature for
the one-chain structure we investigated the height of the first
and second peak of the pair correlation function as function of
temperature (see Fig. 20) in order to look for a structure or an
anomalous jump (as found in Ref. \cite{vsch}) that could identify
the critical temperature. As is apparent from Fig. 20 we do not
find any abrupt changes. The first and second peak as a function
of temperature can be fitted by : $g_i = \alpha (T/T_0)^{-\beta}$,
where $i=\{1$ or $2\}$ denotes the peak (see the curves in Fig.
20).

The values of ($\alpha$,$\beta$) are (2.922, 0.274) for $i=1$ and
(1.895, 0.216) for $i=2$ when $\widetilde{n}_e =0.2$, (9.320,
0.433) for $i=1$ and (7.345, 0.466) for $i=2$ when
$\widetilde{n}_e =0.5$, (14.788, 0.473) for $i=1$ and (11.552,
0.501) for $i=2$ when $\widetilde{n}_e =0.8$ and in each case the
error is less than $1\%$.

From the study of the pair correlation function we conclude that
at moderate ($\tilde{n}_e \leq 0.2$ for $\kappa =1$) densities,
the chain is melted at arbitrarily weak temperature. For higher
densities the chain retains correlations up to higher values of
the temperature but these values are less than those obtained by
the MLC.

We noticed from the high density regime ($\widetilde{n}_e>1$),
where we reach the multichain structure, that another
semi-empirical criterion can be formulated using the pair
correlation function. If we consider the ratio of the height of
the fifth peak in $g(x)$ above 1 ($H_5-1$) to the height of the
first peak above 1 ($H_1-1$) at those densities where the {\it
maximum} melting temperature is obtained for the two, three and
four chain structures, melting occurs when :
\begin{equation}
\frac{H_5-1}{H_1-1}(T_m) \approx 0.15
\end{equation}
\noindent Employing this criterion (termed as pair correlation
function criterion or PCFC) we obtain the results of Fig. 21, were
we present both the relevant temperatures obtained by MLC and
PCFC.

It is worth noticing that this criterion does not work well at
temperatures close to the structural transitions. The reason is
that although particles ``jump'' to new sites in order to attain
the new positions, the pair correlation function still measures
correlations at certain distances and, most importantly, the
height of the first peak is substantially reduced which
artificially enhances the ratio (25).

Thus the value of $15 \%$ which works far from the structural
transitions is too high for the regime close to the structural
transitions. It is therefore evident that the two criteria can
work in a complementary manner.

\section{DISCUSSION AND CONCLUSIONS}

The structural phase transitions and the melting can be studied
experimentally using parabolically confined colloidal particles or
dusty plasmas in the case of a screened Coulomb inter-particle
interaction. Another important experimental system are electrons
floating on liquid helium where it is possible to achieve
relatively narrow Q1D channels on very stable suspended helium
films over structured substrates \cite{sokolov}. Assuming a
semicircular profile of the liquid surface across the channels
then the confining potential is parabolic near the bottom with
$\omega_0 = (e E_{\perp}^*/mR)$ \cite{sokolov} where $E_{\perp}^*$
is the effective holding electric field in the case of the
substrate and R the radius of the semicircular profile. Assuming a
radius of approximately 5 $\mu m$, a typical value for
$E_{\perp}^* \approx 10 kV/cm$ then $\omega_0$ $\approx 10^{11}$
Hz. This in turn produces a $T_0 \approx 60 K$. The melting
temperatures which have been obtained in the present work are of
order $10^{-2} \times T_0$ which results in a melting temperature
$\approx 0.5-1 K$, a temperature range which is routinely achieved
in such experiments. Assuming an interelectron distance of
approximately 0.1 - 1 $\mu m$ leads to a dimensionless linear
density ${\tilde{n}}_e \approx (0.5  - 3) l$, where $l$ is the
number of chains. The dilute limit gives the same ${\tilde{n}}_e$
as the one investigated in the present work.

Another issue connected with melting, which deserves interest, is
the appearance of topological defects so that a KTHNY \cite{kthny}
scenario of melting is possible. In
Ref.\cite{koulakov,bedanov,kong} this question was considered in
the case of a circular confining potential with a finite number of
particles. In the case of short range interactions the defects are
pushed to the surface due to the large price for elastic
deformations while in the Coulomb case shear and Young moduli are
relatively small \cite{koulakov}.

Moreover, because of the incommensurability of the circle with the
hexagonal Wigner crystal the defects do not reside exactly at the
borders but in a zone few lattice spacings inside the crystal.
Therefore three different regimes with different melting
temperatures can be detected \cite{kong}. In our case there is no
such incommensurability and the edges can accommodate the defects
easily. This has also been discussed in the case of a quantum Hall
bar by Nazarov in Ref.\cite{nazarov}.

In conclusion, we investigated the structural, dynamical
properties and melting of a classical quasi-1D system of particles
interacting through a Yukawa-type potential in the range from
Coulomb to very short range interaction in the case where the
confinement is modelled by an external parabolic potential. The
structural transitions are of first (primarily) and second order.
The normal modes of the system were calculated in the presence of
a perpendicular magnetic field. In certain regions of the
parameter space, there is evidence that melting starts first in
the unconfined direction before the system melts along the chain
direction. Furthermore, we found that $T_{m}$ shows a reentrant
behavior as a function of the density of the system and  a regime
of frustration around each point of structural transition can be
identified. In the case of the single chain structure, we device a
new criterion in order to take into account the correlations at
different temperatures. The present study is suitable to describe
colloidal particles, dusty plasmas and electrons floating on
liquid helium.

\section*{ACKNOWLEDGMENTS}

This work was supported in part by the European Community's Human
Potential Programme under contract HPRN-CT-2000-00157 ``Surface
Electrons'', the Flemish Science Foundation (FWO-Vl), IUAP-VI and
the GOA.

\section{APPENDIX A}

The expressions for the energy in the configurations beyond the
two-chain structure are presented below. All the distances are in
units of the interchain distance $a$ between adjacent particles.

For the three-chain structure :
\begin{eqnarray}
\nonumber E_3 = \frac{\widetilde{n}_e}{3} \sum_{m=1}^{\infty}
exp(-3m\kappa /\widetilde{n}_e)/m + \frac{4
\widetilde{n}_e}{9}\sum_{m=1}^{\infty} \frac{exp(-3\kappa
\sqrt{(m-1/2)^2+{c_3}^2}/\widetilde{n}_e)}
{\sqrt{(m-1/2)^2+{c_3}^2}} \\
+ \frac{2 \widetilde{n}_e}{9} \sum_{m=1}^{\infty}
\frac{exp(-3\kappa \sqrt{m^2+4 {c_3}^2} /\widetilde{n}_e)}
{\sqrt{m^2+4 {c_3}^2)}} + 6 \frac{{c_3}^2}{{\widetilde{n}_e}^2}+
\frac{\widetilde{n}_e exp(-6 c_3 \kappa /\widetilde{n}_e)}{18
c_3},
\end{eqnarray}

\noindent where the intrachain distance $c_3$ is a variational
parameter.

For the four-chain structure :

\begin{eqnarray}
\nonumber E_4 &=& \frac{\widetilde{n}_e}{4}\sum_{m=1}^l
\frac{exp(-4 \kappa m/\widetilde{n}_e)}{m}+
\frac{\widetilde{n}_e}{4} \sum_{m=1}^{\infty} \frac{exp(-4 \kappa
\sqrt{(m-1/2)^2+(c_4-f_4)^2}/ \widetilde{n}_e)}
{\sqrt{(m-1/2)^2} + (c_4-f_4)^2}  \\
\nonumber &+& \frac{\widetilde{n}_e}{4} \sum_{m=1}^{\infty}
\frac{exp(-4 \kappa \sqrt{(m^2+(c_4+f_4)^2}/\widetilde{n}_e)}
{\sqrt{m^2+(c_4+f_4)^2}} +\frac{\widetilde{n}_e}{8}
\sum_{m=1}^{\infty} \frac{exp(-4 \kappa \sqrt{(m-1/2)^2+4
{f_4}^2}/\widetilde{n}_e)}
{\sqrt{(m-1/2)^2+4 {f_4}^2}} \\
\nonumber &+&\frac{\widetilde{n}_e}{8} \sum_{m=1}^{\infty}
\frac{exp(-4 \kappa \sqrt{(m-1/2)^2+4 {c_4}^2}/\widetilde{n}_e)}
{\sqrt{(m-1/2)^2+4 {c_4}^2}}
+\frac{8 {c_4}^2}{{\widetilde{n}_e}^2}+8 {f_4}^2/{\widetilde{n}_e}^2 \\
&+&\frac{\widetilde{n}_e}{8} \frac{exp[- 4 (c_4+f_4) \kappa
/\widetilde{n}_e]}{c_4+f_4} ,
\end{eqnarray}

\noindent where $c_4$ is the distance of an inner chain and $f_4$
is the distance of an outer chain from the middle of the
structure. These distances are two variational parameters which
have to be optimized numerically.

For the five-chain structure :

\begin{eqnarray}
\nonumber E_5&=&\frac{10 {c_5}^2}{{\widetilde{n}_e}^2} + \frac{10
{f_5}^2}{{\widetilde{n}_e}^2} + \frac{\widetilde{n}_e}{5}
\sum_{m=1}^{\infty} \frac{exp(-5 \kappa m/\widetilde{n}_e)}{m}
+\frac{4 \widetilde{n}_e}{25} \sum_{m=1}^{\infty} \frac{exp(-5
\kappa \sqrt{(m-1/2)^2+(c_5-f_5)^2}/\widetilde{n}_e)}
{\sqrt{(m-1/2)^2+(c_5-f_5)^2}} \\
\nonumber &+&\frac{4 \widetilde{n}_e}{25} \sum_{m=1}^{\infty}
\frac{exp(-5 \kappa \sqrt{m^2+{c_5}^2}/\widetilde{n}_e)}
{\sqrt{m^2+{c_5}^2}} +\frac{2 \widetilde{n}_e}{25} \frac{exp(-5
c_5 \kappa /\widetilde{n}_e)}{c_5} + \frac{4 \widetilde{n}_e}{25}
\sum_{m=1}^{\infty} \frac{exp(-5 \kappa
\sqrt{(m-1/2)^2+(c_5+f_5)^2}/\widetilde{n}_e)}
{\sqrt{(m-1/2)^2+(c_5+f_5)^2}} \\
\nonumber &+&\frac{2 \widetilde{n}_e}{25} \sum_{m=1}^{\infty}
\frac{exp(-5 \kappa \sqrt{m^2+4 {c_5}^2}/\widetilde{n}_e)}
{\sqrt{m^2+4 {c_5}^2}} +\frac{\widetilde{n}_e}{50} \frac{exp(-10
c_5 \kappa /\widetilde{n}_e)}{c_5} +\frac{4 \widetilde{n}_e}{25}
\sum_{m=1}^{\infty} \frac{exp(-5 \kappa
\sqrt{(m-1/2)^2+{f_5}^2}/\widetilde{n}_e)}
{\sqrt{(m-1/2)^2+{f_5}^2}} \\
&+&\frac{2 \widetilde{n}_e}{25} \sum_{m=1}^{\infty} \frac{exp(-5
\kappa \sqrt{m^2+4{f5}^2)/ \widetilde{n}_e})} {\sqrt{m^2+4
{f_5}^2}}+ \frac{\widetilde{n}_e exp(-10 f_5 \kappa
/\widetilde{n}_e)}{50 f_5} ,
\end{eqnarray}
\noindent with the variational parameters $c_5$ and $f_5$, which
are the distance of an inner chain and outer chain respectively
from the middle of the structure.

For the six-chain structure :

\begin{eqnarray}
\nonumber E_6&=&\frac{12 {f_6}^2}{{\widetilde{n}_e}^2} +\frac{12
{g_6}^2}{{\widetilde{n}_e}^2} +\frac{12
{h_6}^2}{{\widetilde{n}_e}^2}+ \frac{\widetilde{n}_e}{6}
\sum_{m=1}^{\infty} \frac{exp(-6 \kappa m/6)}{m}+
\frac{\widetilde{n}_e}{18} \sum _{m=1}^{\infty} (\frac{exp(-6
\kappa \sqrt{(m-1/2)^2+4 {h_6}^2}/\widetilde{n}_e)}{\sqrt{(m-1/2)^2+4 {h_6}^2)}}\\
\nonumber &+&\frac{\widetilde{n}_e}{18} \sum_{m=1}^{\infty}
(\frac{exp(-6 \kappa \sqrt{(m-1/2)^2+4 {f_6}^2}/\widetilde{n}_e)}
{\sqrt{(m-1/2)^2+4 {h_6}^2)}} +\frac{\widetilde{n}_e}{18}
\sum_{m=1}^{\infty} (\frac{exp(-6 \kappa \sqrt{(m-1/2)^2+4
{g_6}^2}/\widetilde{n}_e)}
{\sqrt{(m-1/2)^2+4 {g_6}^2}}) \\
\nonumber &+&\frac{\widetilde{n}_e}{9} \sum_{m=1}^{\infty}
\frac{exp(-6 \kappa \sqrt{(m-1/2)^2+(h_6-g_6)^2}/\widetilde{n}_e)}
{\sqrt{(m-1/2)^2+(h_6-g_6)^2}} +\frac{\widetilde{n}_e}{9}
\sum_{m=1}^{\infty} \frac{exp(-6 \kappa
\sqrt{(m-1/2)^2+(h_6+f_6)^2}/\widetilde{n}_e)}
{\sqrt{(m-1/2)^2+(h_6+f_6)^2)}} \\
\nonumber &+&\frac{\widetilde{n}_e}{9} \sum_{m=1}^{\infty}
\frac{exp(-6 \kappa \sqrt{(m-1/2)^2+(g_6-f_6)^2}/\widetilde{n}_e)}
{\sqrt{(m-1/2)^2 +(g_6-f_6)^2)}} +\frac{\widetilde{n}_e}{9}
\sum_{m=1}^{\infty} \frac{exp(-6 \kappa
\sqrt{m^2+(h_6-f_6)^2)}/\widetilde{n}_e)}
{\sqrt{m^2+(h_6-f_6)^2}} \\
\nonumber &+&\frac{\widetilde{n}_e}{18} \frac{exp(-6 |h_6-f_6|
\kappa /\widetilde{n}_e)}{{|h_6-f_6|}} +\frac{\widetilde{n}_e}{9}
\sum_{m=1}^{\infty} \frac{exp(-6 \kappa
\sqrt{m^2+(h_6+g_6)^2}/\widetilde{n}_e)} {\sqrt{m^2+(h_6+g_6)^2}}
+\frac{\widetilde{n}_e}{18} \frac{exp[-6 (h_6+g_6) \kappa
/\widetilde{n}_e]}
{{(h_6+g_6)}} \\
&+& \frac{\widetilde{n}_e}{9} \sum_{m=1}^{\infty}
\frac{exp(-6\kappa \sqrt{m^2+(g_6+f_6)^2}/\widetilde{n}_e)}
{\sqrt{m^2+(g_6+f_6)^2}} +\frac{\widetilde{n}_e}{18} \frac{exp[-6
(g_6+f_6)  \kappa /\widetilde{n}_e]}{(g_6+f_6)} ,
\end{eqnarray}
\noindent with $f_6$, $g_6$ and $h_6$ the three chain distances
from the middle of the crystal, starting from the inner one which
are the variational parameters.

\section{APPENDIX B}

We present the matrix $\omega^{2} {\bf I} -{\bf D}$ where ${\bf
I}$ is the unit matrix and ${\bf D}$ is the dynamical matrix, for
the calculation of the normal modes for the three-chain structure.
It reads :

\begin{equation}
\omega^{2} {\bf I} -{\bf D} =
\begin{pmatrix}  {\omega}^2 -A_1 & 0 & -A_3 & 0 & -A_5 & 0 \\
0 & ({\omega}^2 -{\omega_0}^2) - A_2 & 0 & -A_4 & 0 & -A_6 \\
-A_3 & 0 &  {\omega}^2 -A_1 & 0 & -A_3 & 0 \\
0 & -A_4 & 0 &  ({\omega}^2 - {\omega_0}^2) - A_2 & 0 & -A_4 \\
-A_5 & 0 & -A_3 & 0 &  {\omega}^2 - A_1 & 0 \\
0 & -A_6 & 0 & -A_4 & 0 &  ({\omega}^2 - {\omega_0}^2) -
A_2\end{pmatrix} \quad
\end{equation}

\noindent where the parameters are :

\begin{eqnarray}
\nonumber A_1 &=& {\tilde{n}_e}^3 \sum_{j=1}^{\infty} \frac{1}{27
j^3} exp(-3 j \kappa /{\tilde{n}_e}) [2+ 6 j \kappa /\tilde{n}_e +
9 j^2 {\kappa}^2 /{\tilde{n}_e}^2]
(1-cos(\tilde{k} \pi j)) , \\
\nonumber A_2 &=& -{\tilde{n}_e}^3 \sum_{j=1}^{\infty} \frac{1}{27
j^3} exp(-3 j \kappa  /{\tilde{n}_e})
(2 + 3 j \kappa  /\tilde{n}_e ) (1-cos(\tilde{k} \pi j)) , \\
\nonumber A_3 &=& {\tilde{n}_e}^3 \sum_{j=1}^{\infty} \frac{1}{27
\sqrt{(j+1/2)^2+c^2}} exp(-3 \kappa \sqrt{(j+1/2)^2+c^2}/
\tilde{n}_e) [cos(\tilde{k} \pi (j+1/2))-1 ] (j+1/2)^2 \times \\
\nonumber &&[\frac{9 \kappa \sqrt{(j+1/2)^2
+{c_3}^2}}{\tilde{n}_e} + \frac{9 {\kappa}^2 [(j+1/2)^2 +{c_3}^2]}
{{\tilde{n}_e}^2}+ 3] - (1 + \frac{3 \kappa
\sqrt{(j+1/2)^2+c^2}}{\tilde{n}_e})
[(j+1/2)^2 + {c_3}^2] , \\
\nonumber A_4 &=& {\tilde{n}_e}^3 \sum_{j=1}^{\infty} \frac{1}{27
\sqrt{(j+1/2)^2+{c_3}^2}} exp(-3 \kappa \sqrt{(j+1/2)^2+{c_3}^2}/
\tilde{n}_e) [cos(\tilde{k} \pi (j+1/2))-1 ] {c_3}^2 \times \\
\nonumber &&[\frac{9 \kappa \sqrt{(j+1/2)^2
+{c_3}^2}}{\tilde{n}_e} + \frac{9 {\kappa}^2 [(j+1/2)^2 +{c_3}^2]}
{{\tilde{n}_e}^2}+ 3] - (1 + \frac{3 \kappa
\sqrt{(j+1/2)^2+{c_3}^2}}{\tilde{n}_e})
[(j+1/2)^2 + {c_3}^2] , \\
\nonumber A_5 &=& {\tilde{n}_e}^3  \sum_{j=1}^{\infty} \frac{1}{27
\sqrt{j^2+4 {c_3}^2}} exp(-3 \kappa \sqrt{j^2+4 c^2}/\tilde{n}_e)
[cos(\tilde{k} \pi j) - 1]  \times \\
\nonumber &&[ (\frac{9 \kappa \sqrt{j^2 + 4 {c_3}^2}}{\tilde{n}_e}
+ \frac{9 {\kappa}^2 (j^2 + 4 {c_3}^2)}{{\tilde{n}_e}^2} + 3)(j^2
+ 4 {c_3}^2)-(1 + \frac{3 \kappa \sqrt{j^2 + 4 {c_3}^2}}{\tilde{n}_e})(j^2+4 {c_3}^2)] ,\\
\nonumber A_6 &=& {\tilde{n}_e}^3 \sum_{j=1}^{\infty} \frac{1}{27
\sqrt{j^2+4 {c_3}^2}} exp(-3 \kappa \sqrt{j^2+4
{c_3}^2}/\tilde{n}_e)
[cos(\tilde{k} \pi j) - 1]  \times \\
\nonumber &&[ (\frac{9 \kappa \sqrt{j^2 + 4 {c_3}^2}}{\tilde{n}_e}
+ \frac{9 {\kappa}^2 (j^2 + 4 {c_3}^2)}{{\tilde{n}_e}^2} + 3) 4
{c_3}^2 -(1 + \frac{3 \kappa \sqrt{j^2 + 4
{c_3}^2}}{\tilde{n}_e})(j^2+4 {c_3}^2)] ,
\end{eqnarray}

\noindent where $\tilde{k} = a k /\pi$, is the dimensionless wavenumber.

The modes for the single-chain are obtained by the top left part
of the matrix (32) which forms a 2 $\times$ 2 submatrix and
involves the elements $A_1$ and $A_2$ which have exactly the same
form with the substitution ${\tilde{n}_e}/3 \rightarrow
\tilde{n}_e$.

Similarly, the modes for the two-chain structure can be obtained
by the 3 $\times$ 3 submatrix which is included in the top left
part of matrix (32) and involves the elements $A_1$, $A_2$ and
$A_3$ with the substitution ${\tilde{n}_e}/3 \rightarrow
\tilde{n}_e /2$.

\clearpage

\begin{figure}
\begin{center}
\includegraphics[width=10cm]{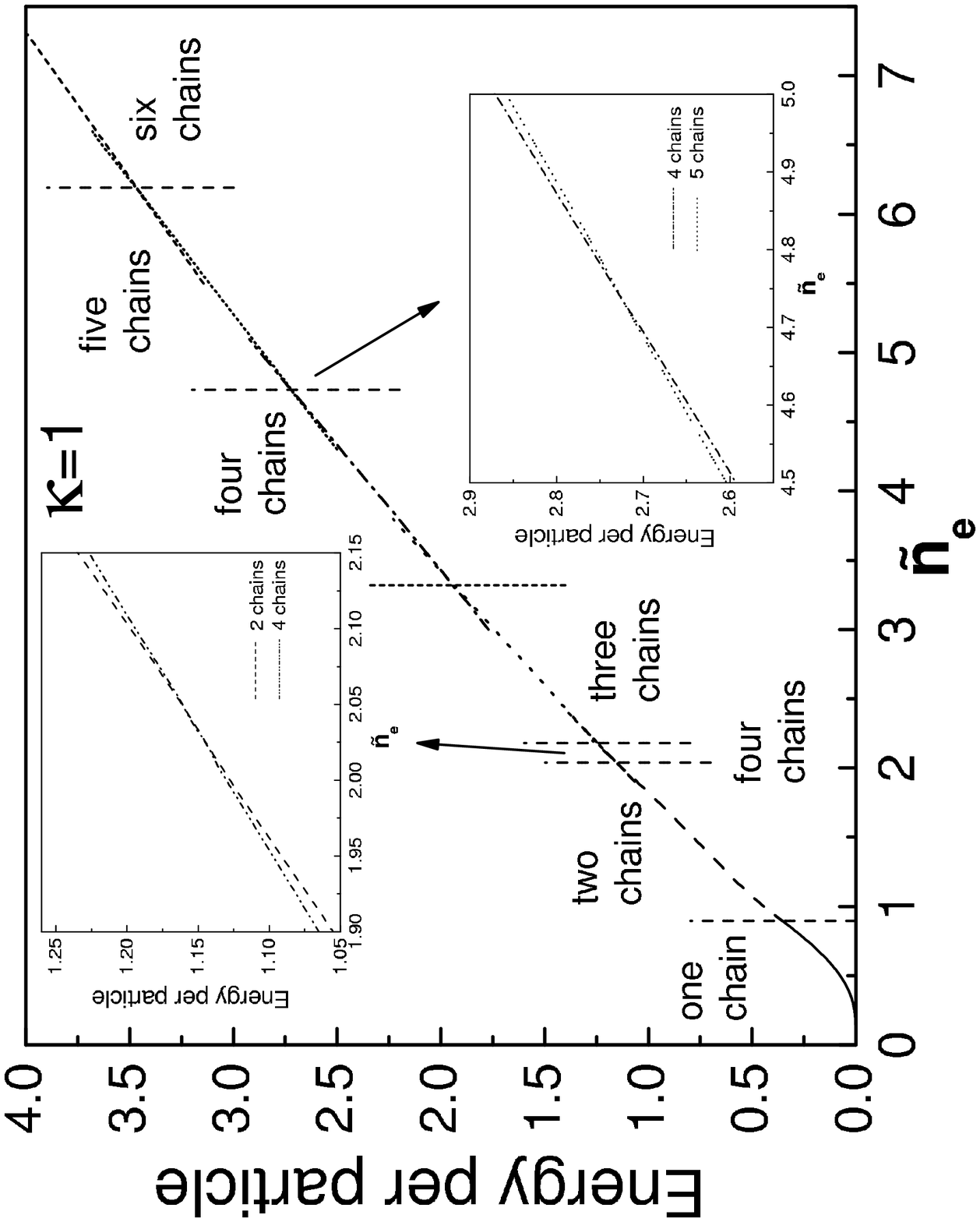}
\caption{The energy per particle as a function of density for
$\kappa$=1.}
\label{figurename}
\end{center}
\end{figure}

\begin{figure}
\begin{center}
\includegraphics[width=10cm]{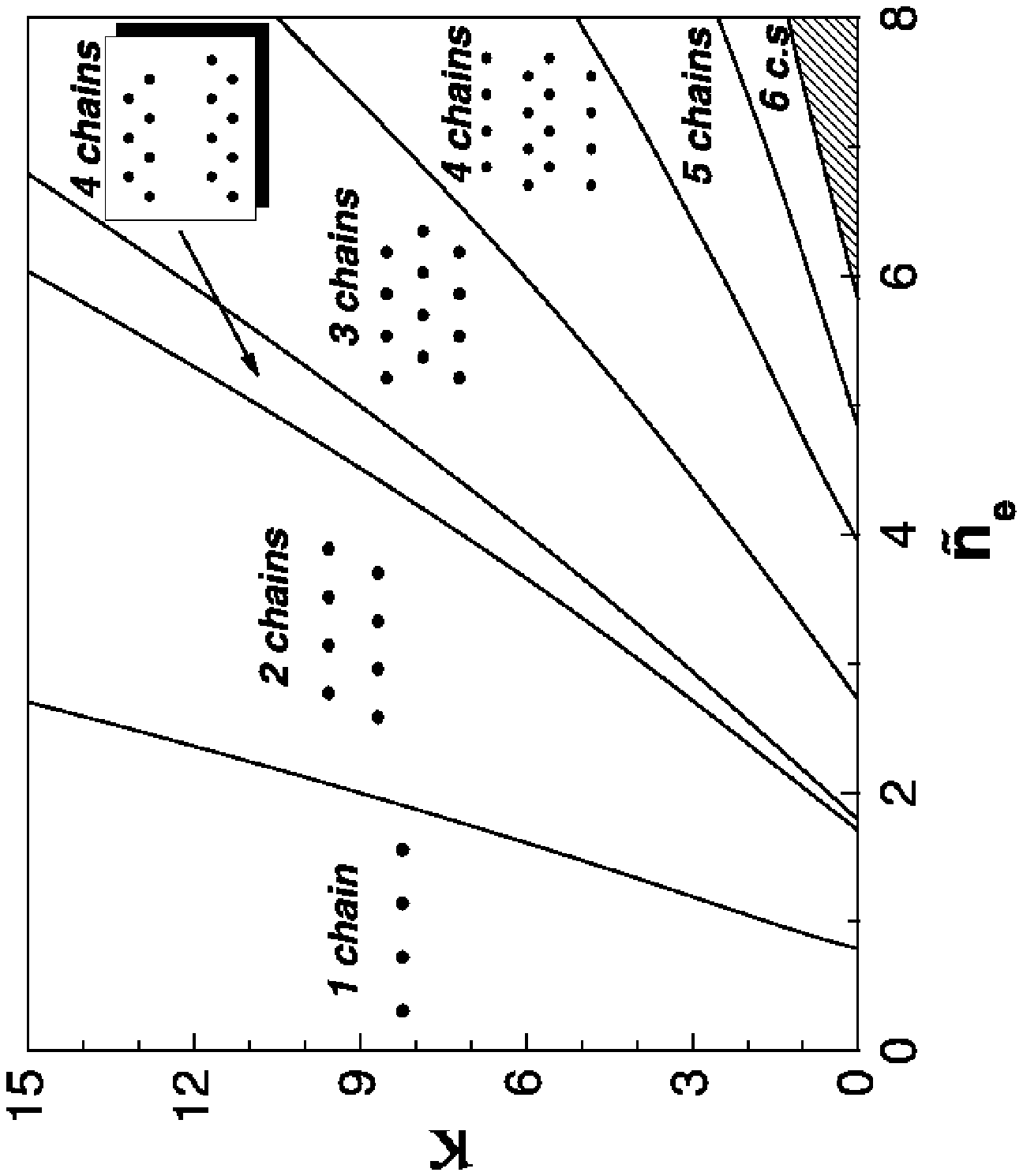}
\caption{The zero temperature structural phase diagram.}
\label{figurename}
\end{center}
\end{figure}

\begin{figure}
\begin{center}
\includegraphics[width=10cm]{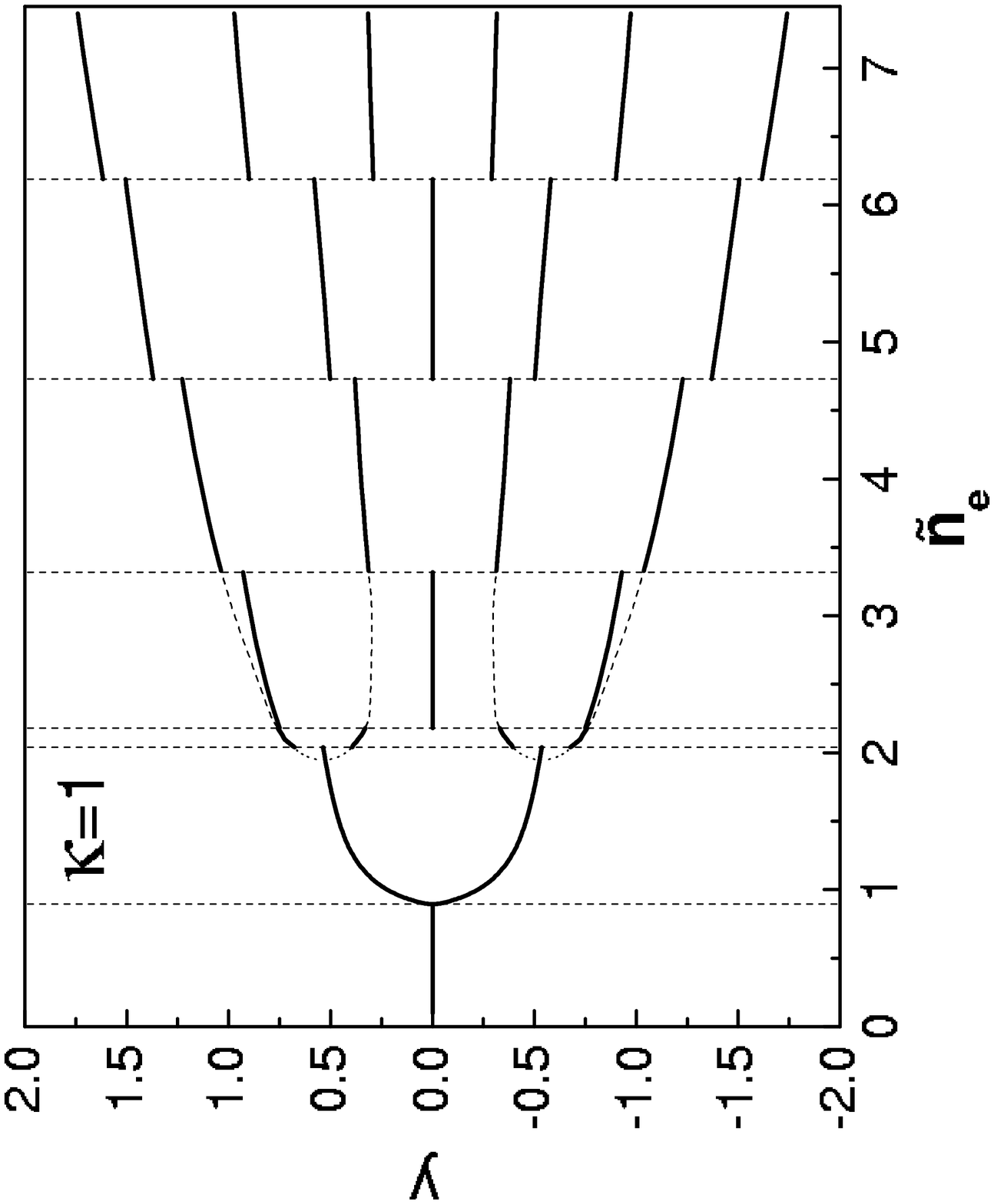}
\caption{The lateral position of the chains in the Wigner crystal
state as a function of the linear density for $\kappa=1$.}
\label{figurename}
\end{center}
\end{figure}

\begin{figure}
\begin{center}
\includegraphics[width=10cm]{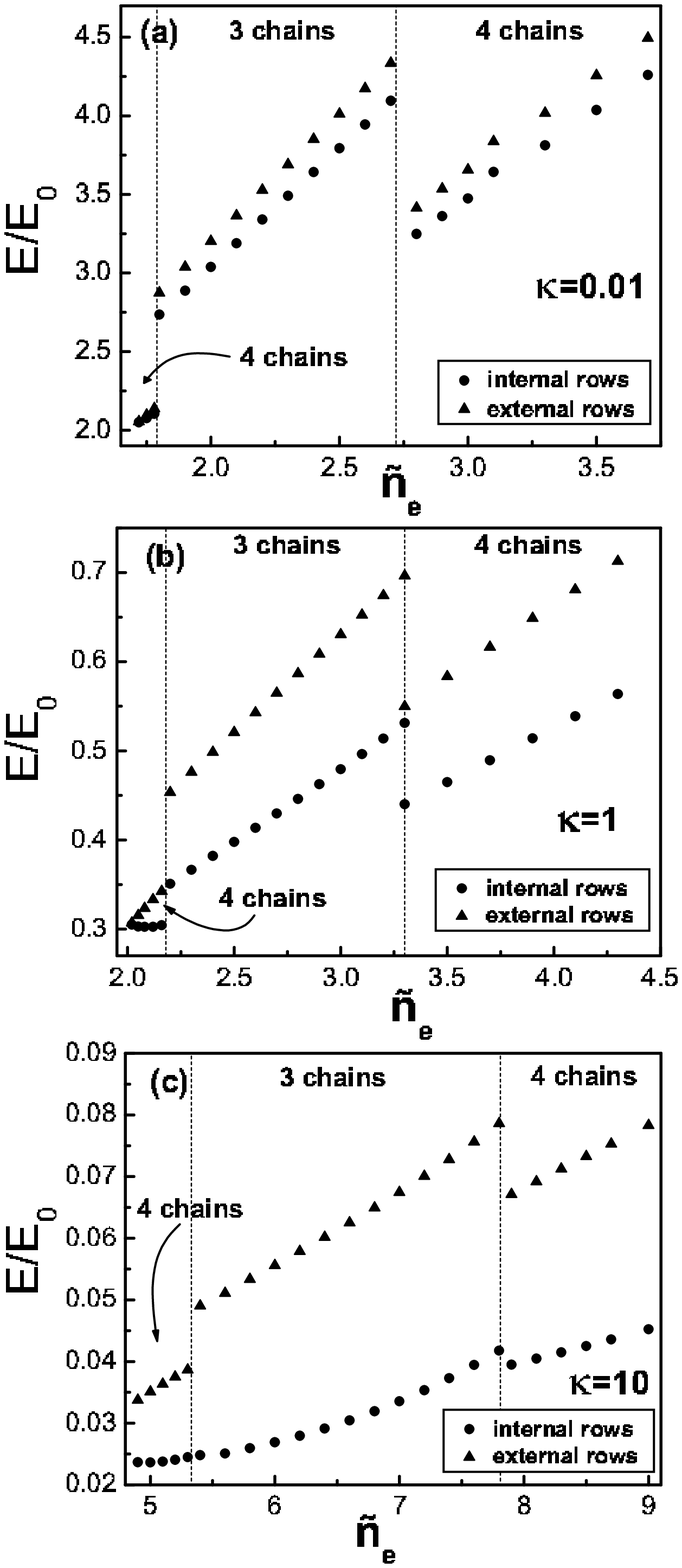}
\caption{The energy per chain at T=0 and (a) $\kappa$=0.01, (b)
$\kappa$=1 and (c)$\kappa$=10. The energy is always higher for the
external chains but as the Coulomb limit ($\kappa \ll 1$) is
approached the difference is diminished and the system behaves
isotropically.} \label{figurename}
\end{center}
\end{figure}

\begin{figure}
\begin{center}
\includegraphics[width=10cm]{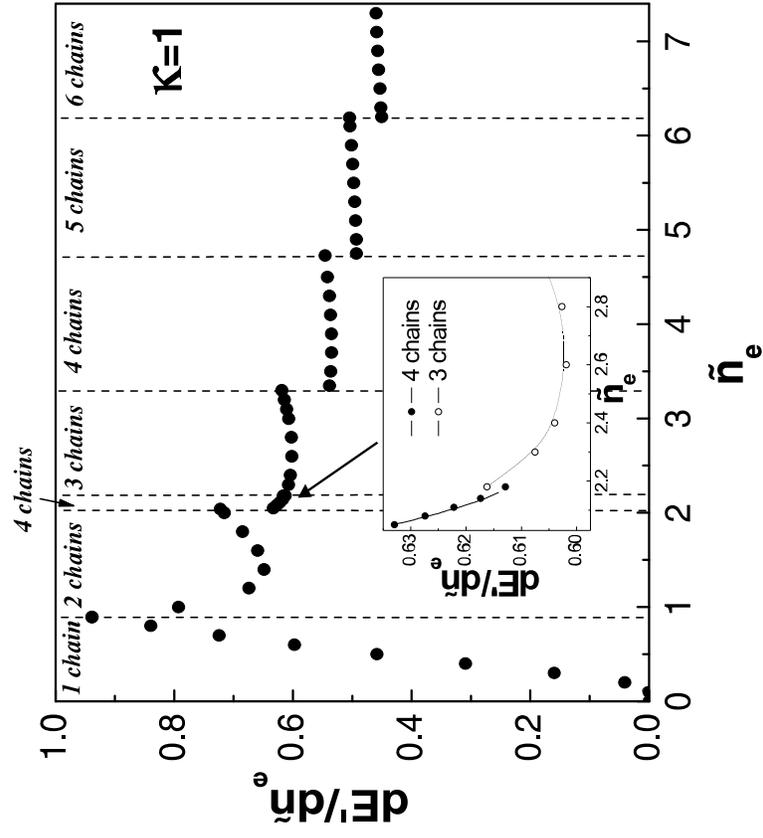}
\caption{The derivative of the energy with respect to the density
for $\kappa =1$. Only the transition from one to two-wires is
continuous (second order) the rest are first order.}
\label{figurename}
\end{center}
\end{figure}

\begin{figure}
\begin{center}
\includegraphics[width=15cm]{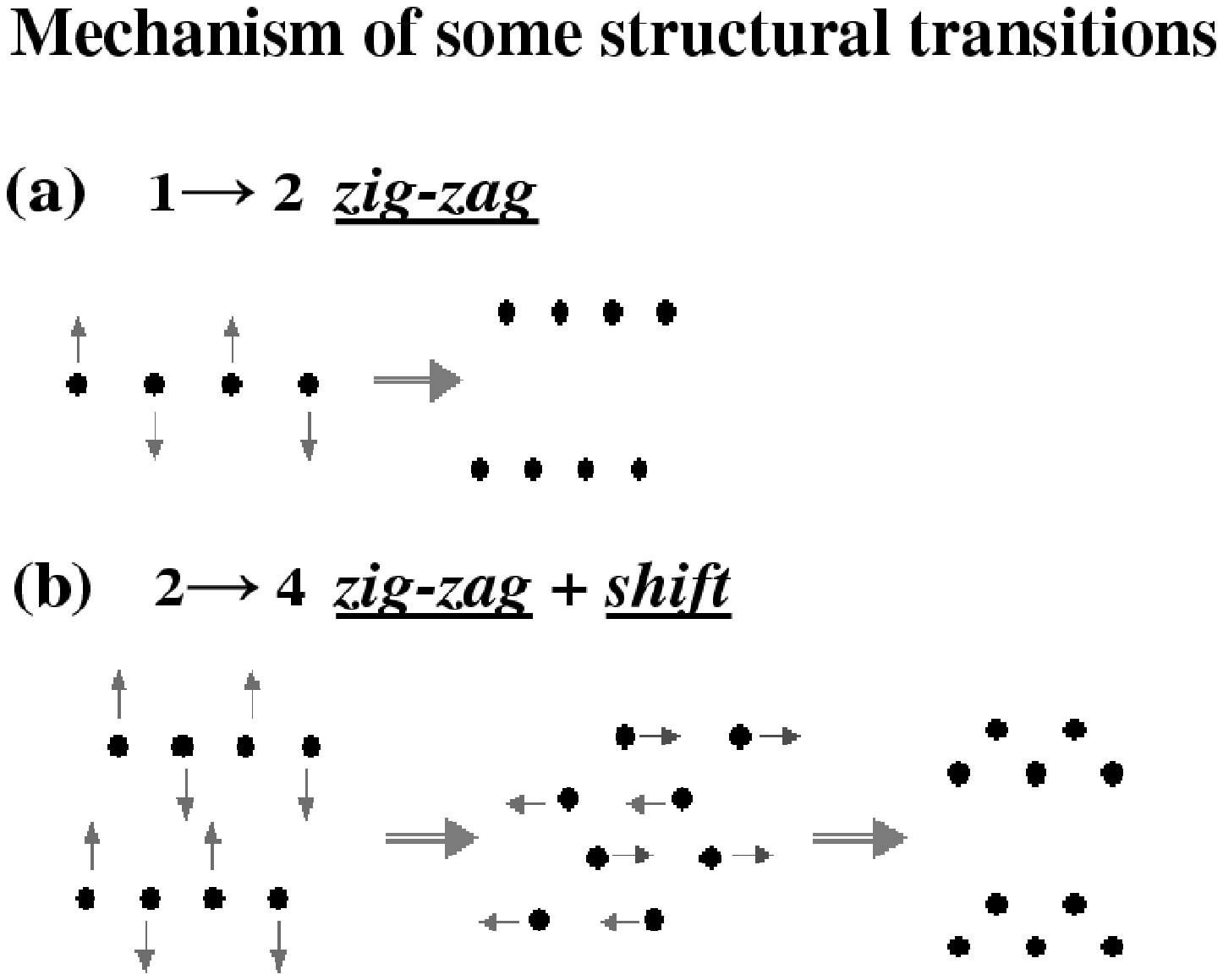}
\caption{The mechanism of the structural transitions
1$\longrightarrow$ 2 chains and 2 $\longrightarrow$ 4 chains.}
\label{figurename}
\end{center}
\end{figure}

\begin{figure}
\begin{center}
\includegraphics[width=15cm]{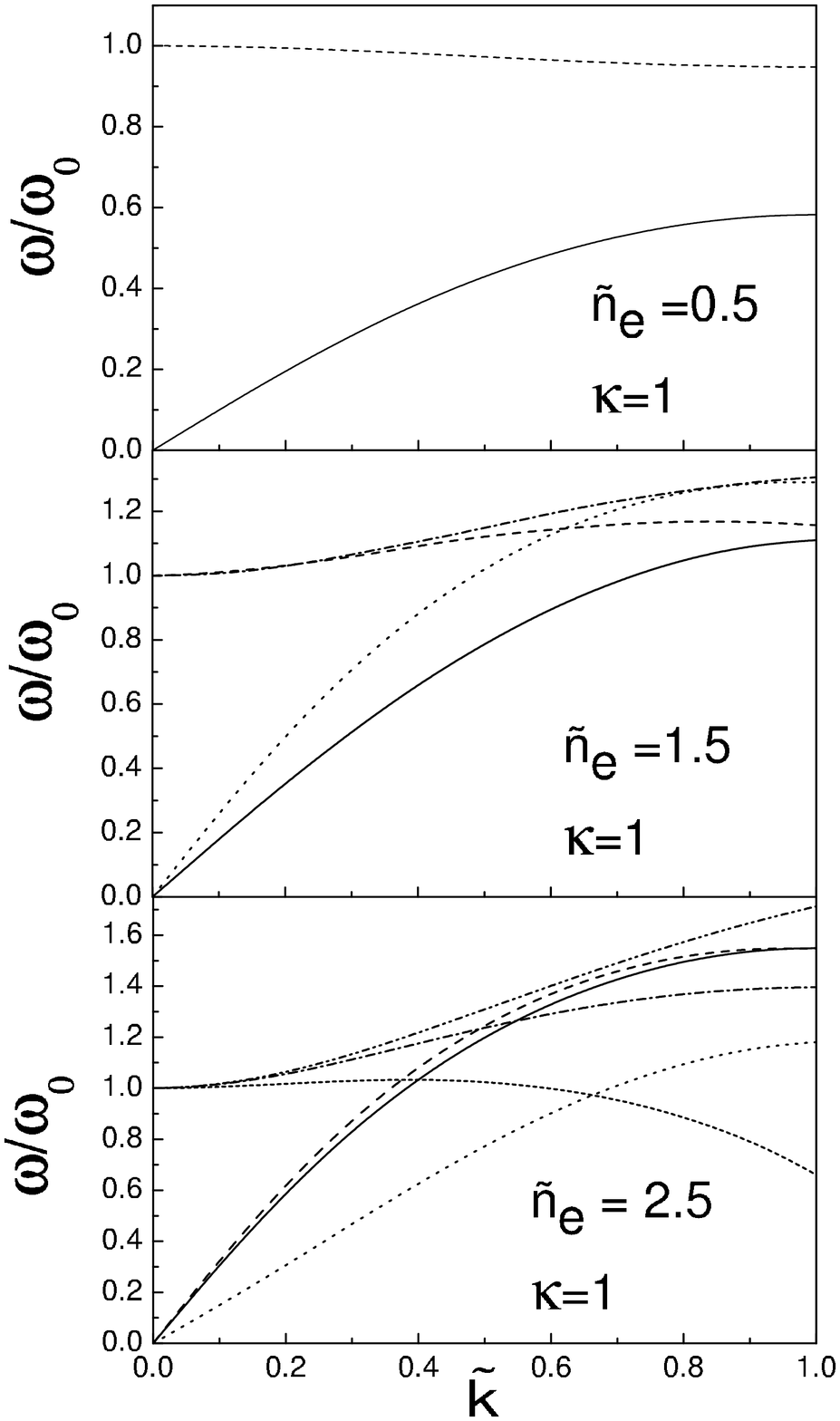}
\caption{The normal modes of the system in the one (a), two (b)
and three (c) chain configuration. The optical and acoustical
branches correspond to motion in the confined and unconfined
direction respectively. The wavelength is in units of $\pi/a$,
where $a$ is the length of the unit cell.} \label{figurename}
\end{center}
\end{figure}

\begin{figure}
\begin{center}
\includegraphics[width=10cm]{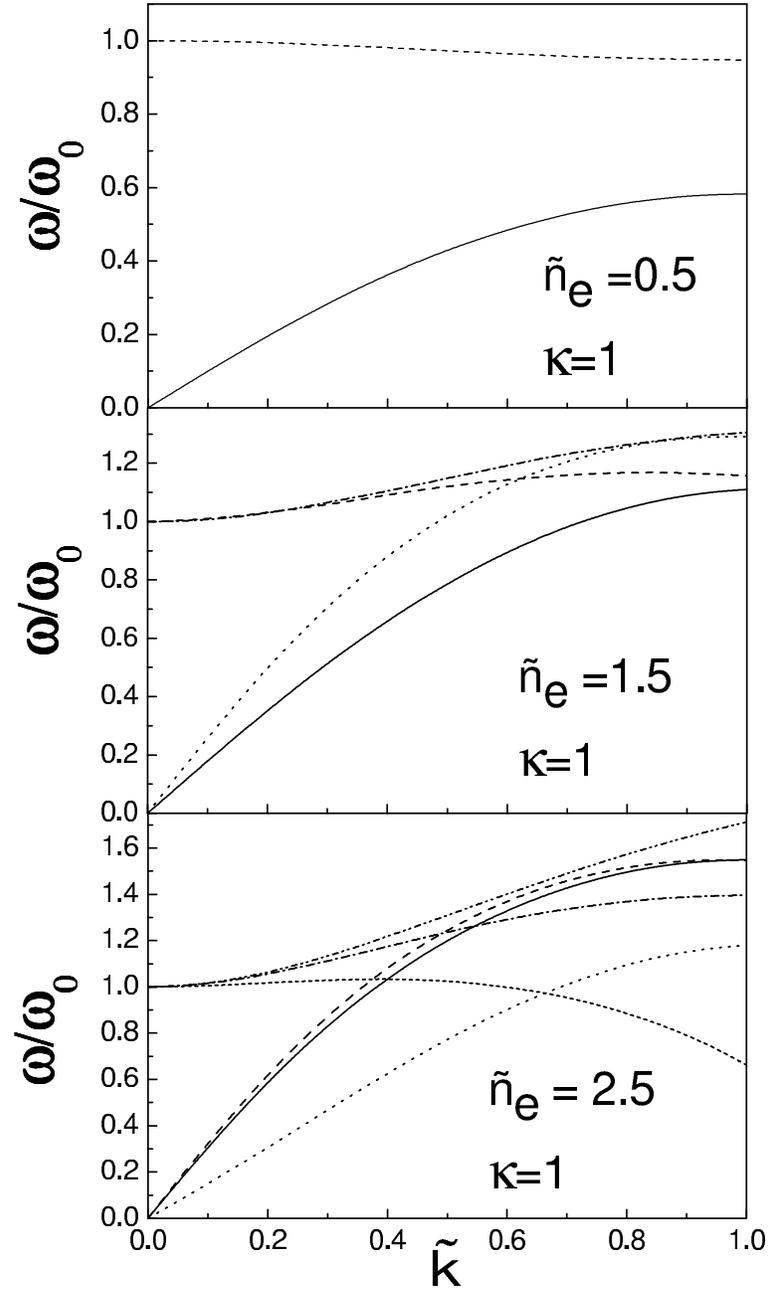}
\caption{The phonon spectrum at the softening of the optical mode
at the structural transition from one to two-chains.}
\label{figurename}
\end{center}
\end{figure}

\begin{figure}
\begin{center}
\includegraphics[width=10cm]{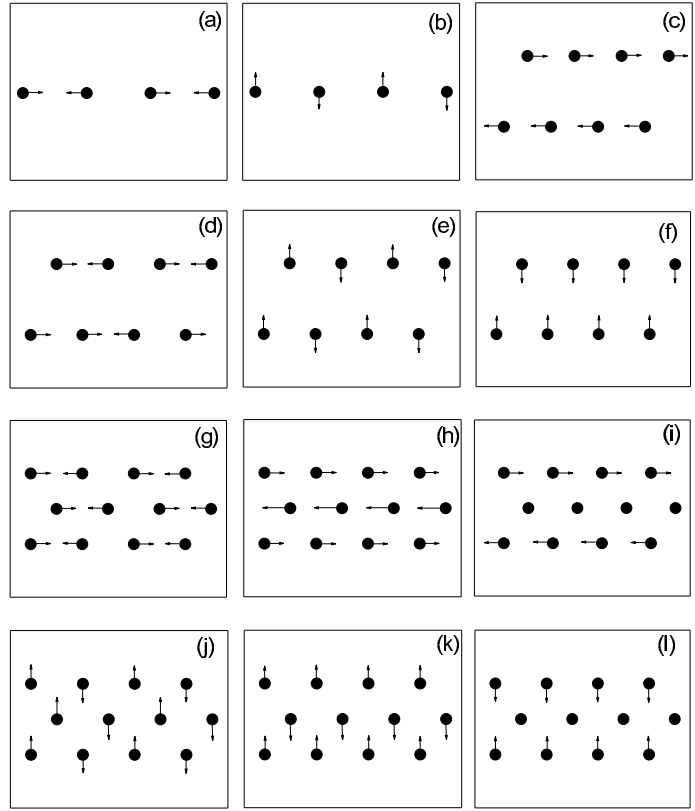}
\caption{The motion of the particles for the one-, two- and three-
chain structure which corresponds to the different
eigenfrequencies.} \label{figurename}
\end{center}
\end{figure}

\begin{figure}
\begin{center}
\includegraphics[height=12cm]{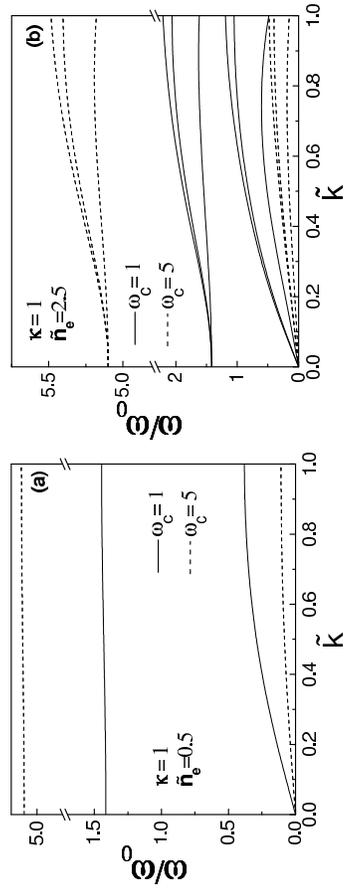}
\caption{Typical dispersion curves for the one (a) and three (b)
chain structures for two different magnetic field values.}
\label{figurename}
\end{center}
\end{figure}

\clearpage

\begin{figure}
\begin{center}
\includegraphics[width=10cm]{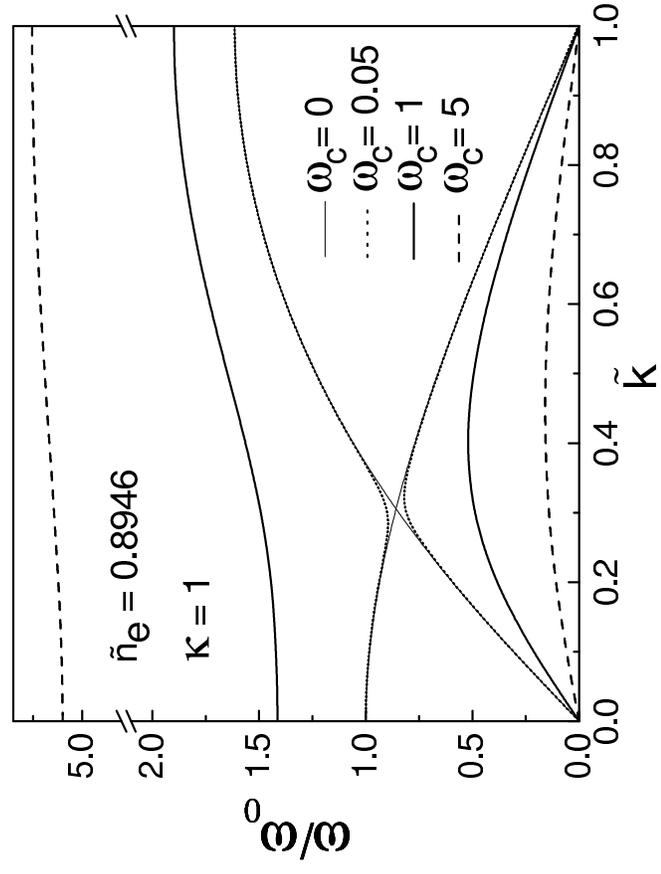}
\caption{The magnetic field dependence of the softening of the
phonon mode at the structural transition from one to two-chains.}
\label{figurename}
\end{center}
\end{figure}

\begin{figure}
\begin{center}
\includegraphics[width=15cm]{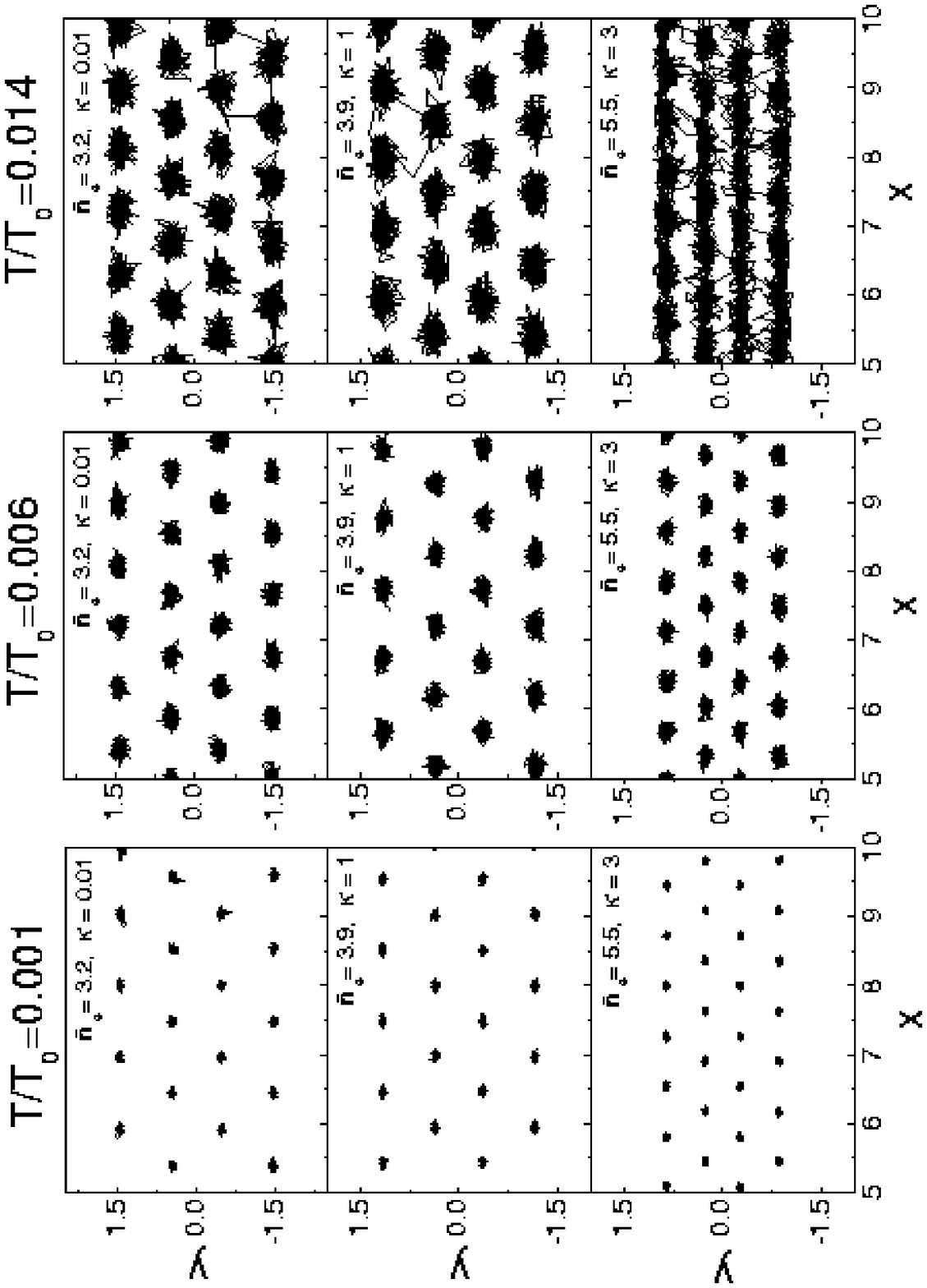}
\caption{Particle trajectories for $3 \times 10^7$ MC steps for
three different temperatures and three different values of the
density at $\kappa$ = 1.} \label{figurename}
\end{center}
\end{figure}

\begin{figure}
\begin{center}
\includegraphics[width=10cm]{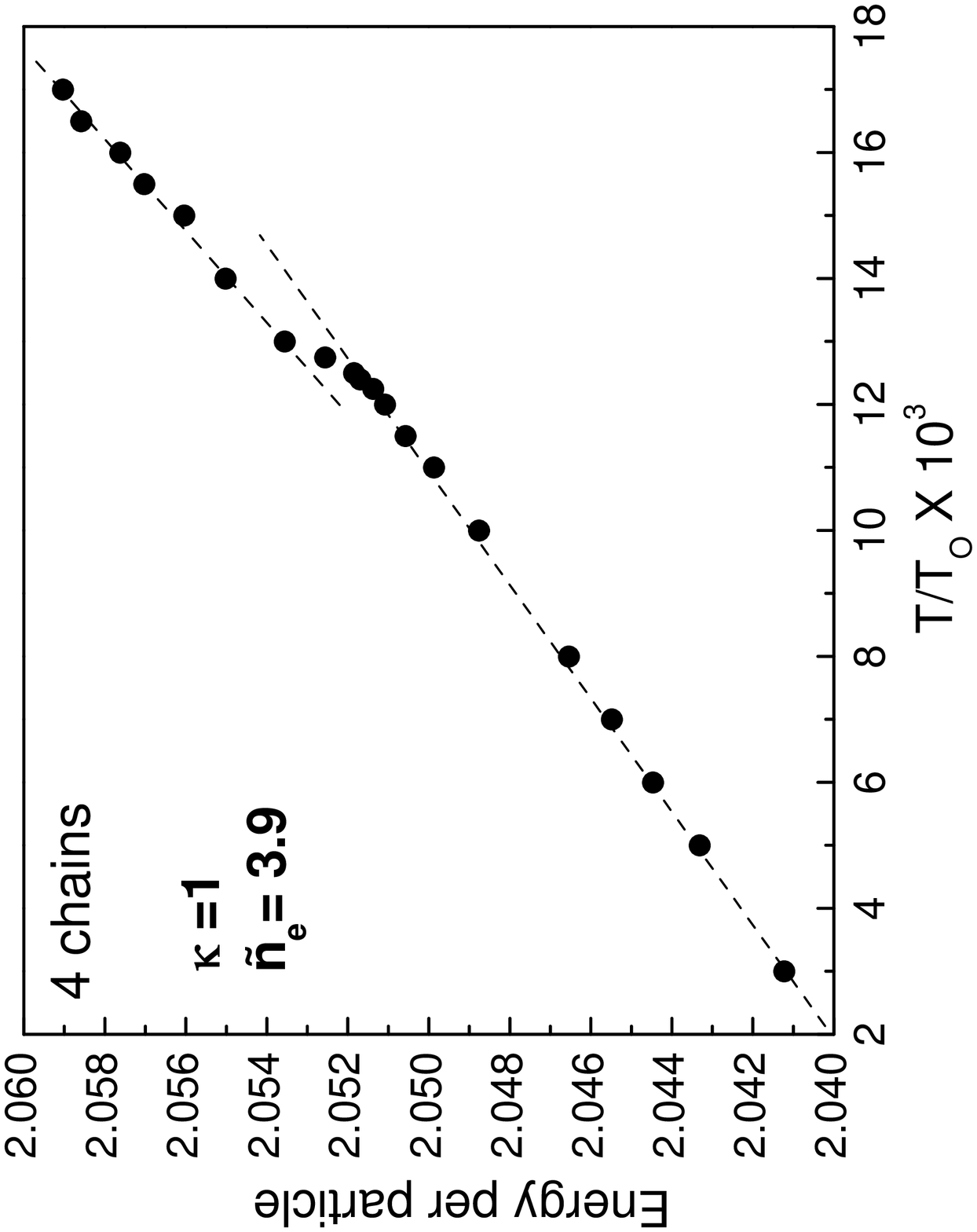}
\caption{The energy per particle as a function of temperature for
the four-chain structure with $\kappa = 1$ and
$\widetilde{n}_e=3.9$. There is a fast increase of the energy at
the melting temperature.} \label{figurename}
\end{center}
\end{figure}

\begin{figure}
\begin{center}
\includegraphics[width=10cm]{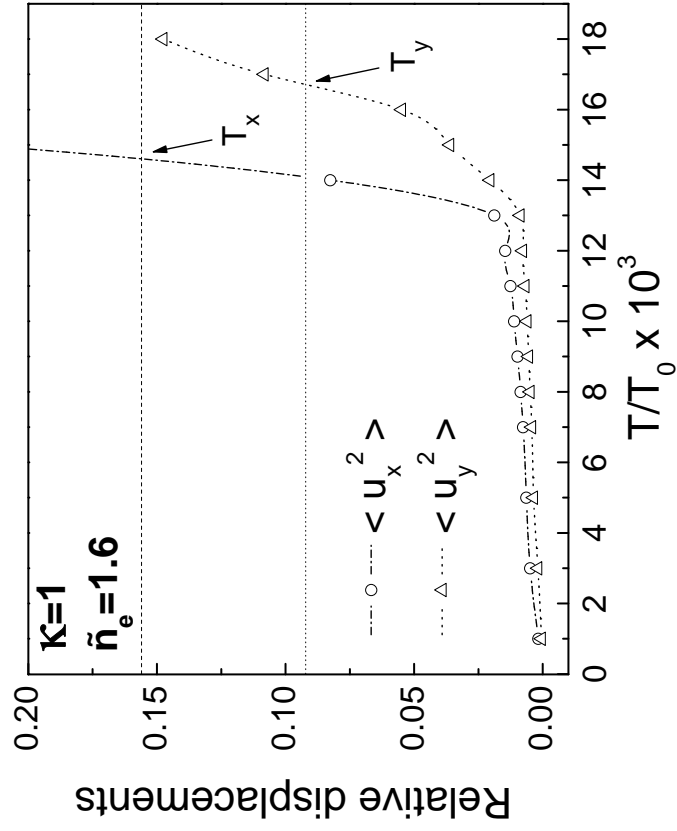}
\caption{The mean square relative displacements for $\kappa$= 1,
$\widetilde{n}_e$= 1.6 in the case of the two-chain configuration.
The dashed horizontal line corresponds to the modified Lindemann
criterion (MLC) in the unconfined direction, while the dotted line
corresponds to the MLC in the confined direction.}
\label{figurename}
\end{center}
\end{figure}

\begin{figure}
\begin{center}
\includegraphics[width=10cm]{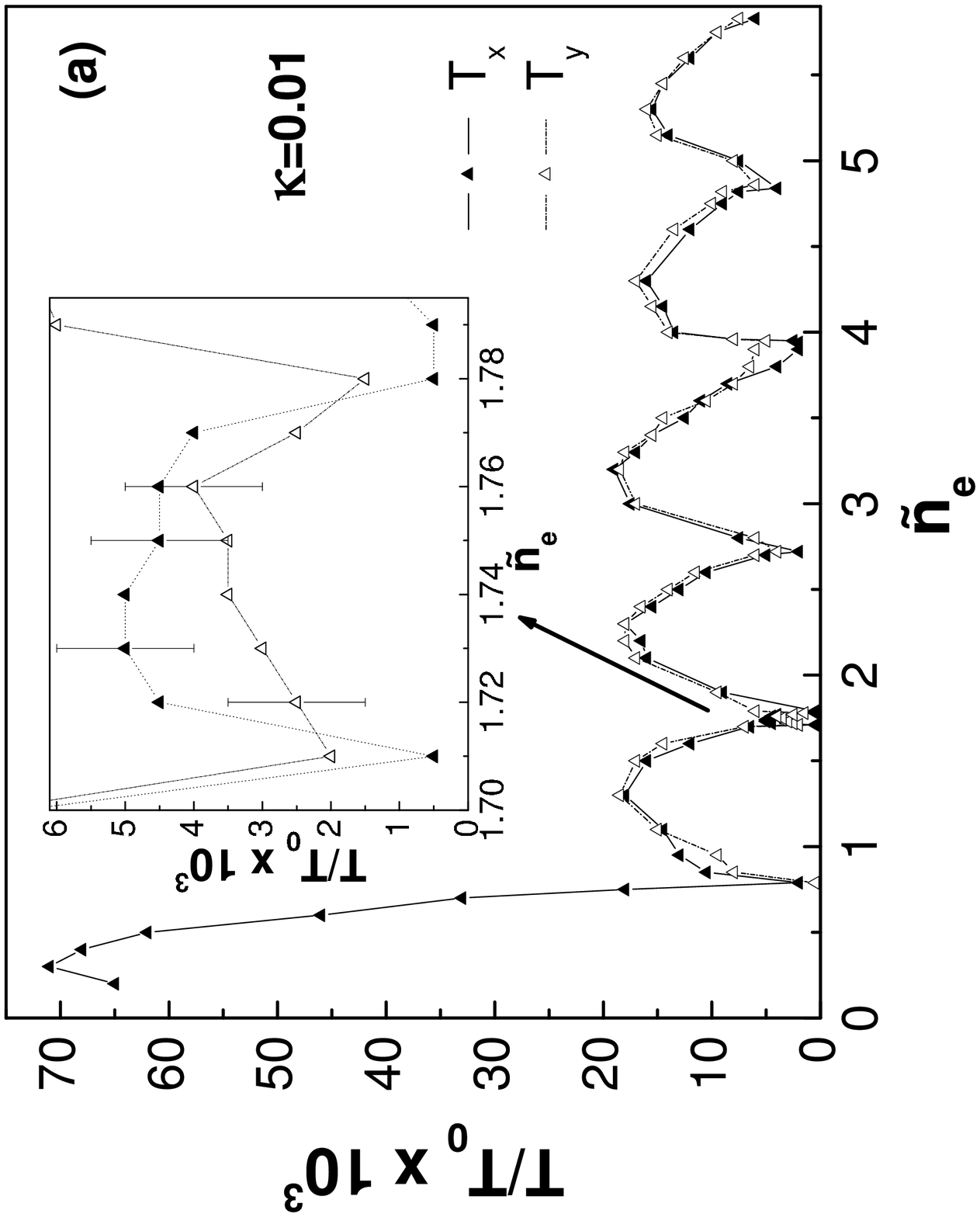}
\includegraphics[width=10cm]{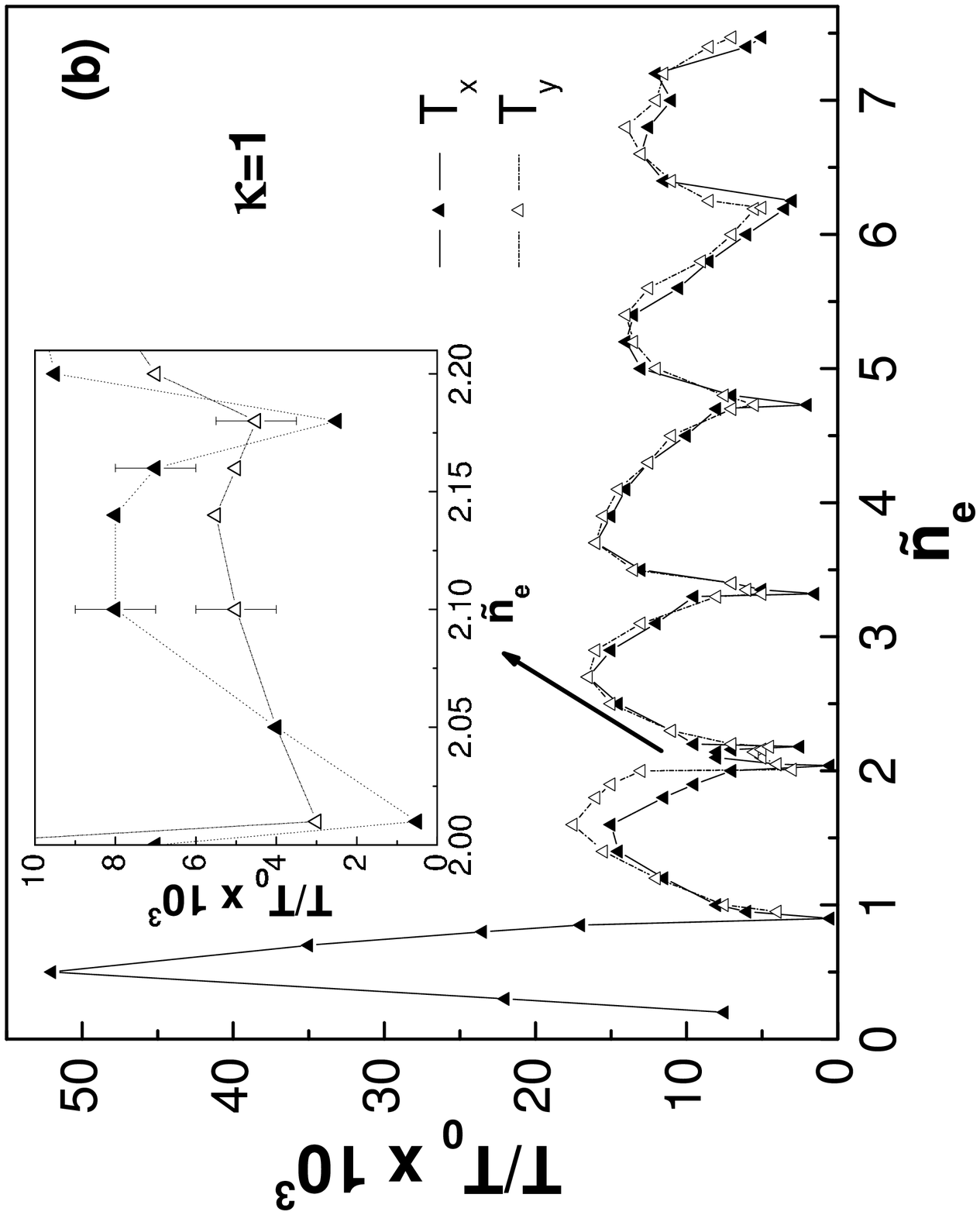}
\includegraphics[width=10cm]{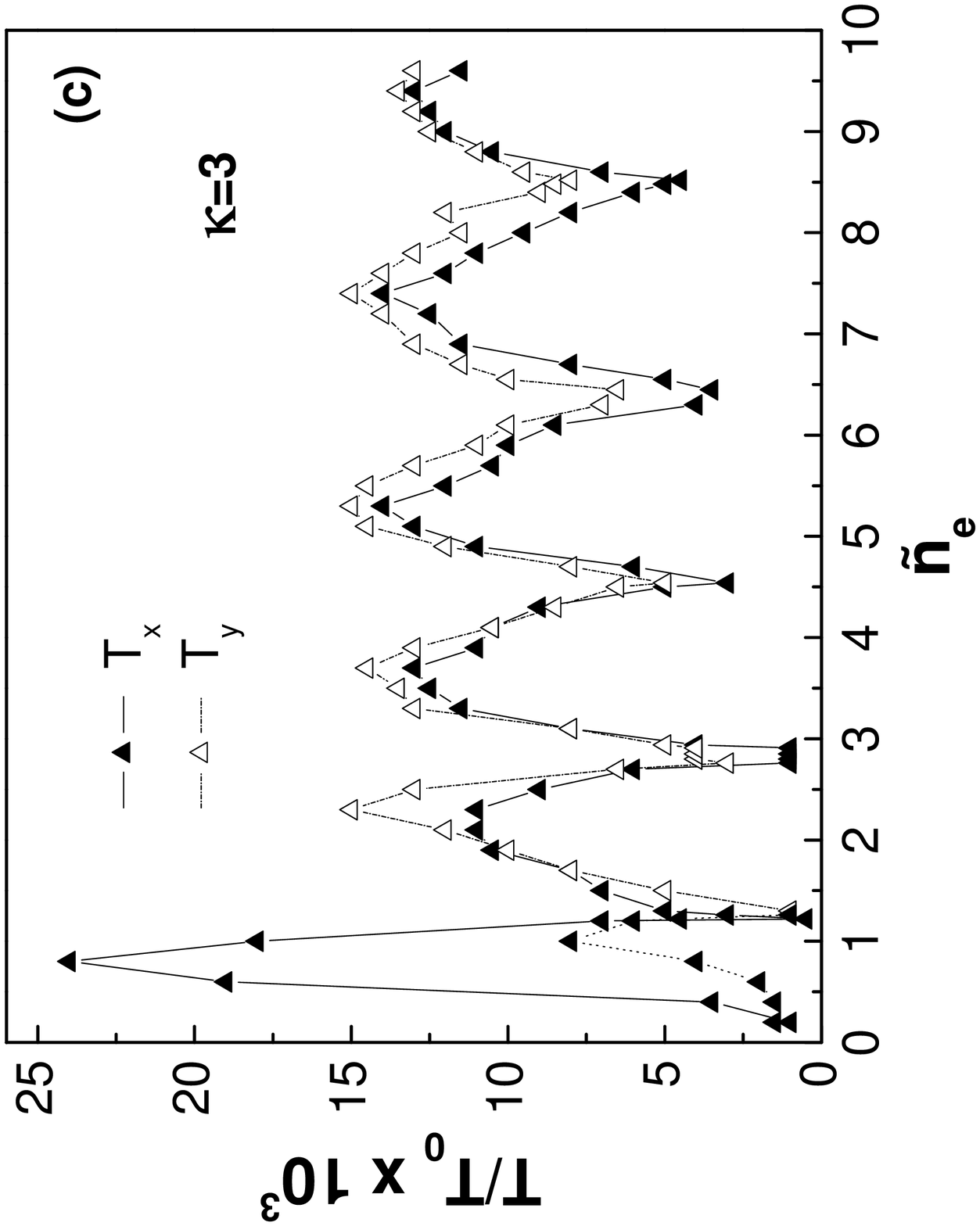}
\caption{Melting temperature as a function of density for: (a)
$\kappa$ = 0.01, (b) $\kappa$=1, and (c) $\kappa=3$. The insets in
(a) and (b) show an enlargement of the four-chain region which is
located between the two and three chain phase.} \label{figurename}
\end{center}
\end{figure}

\begin{figure}
\begin{center}
\includegraphics[width=15cm]{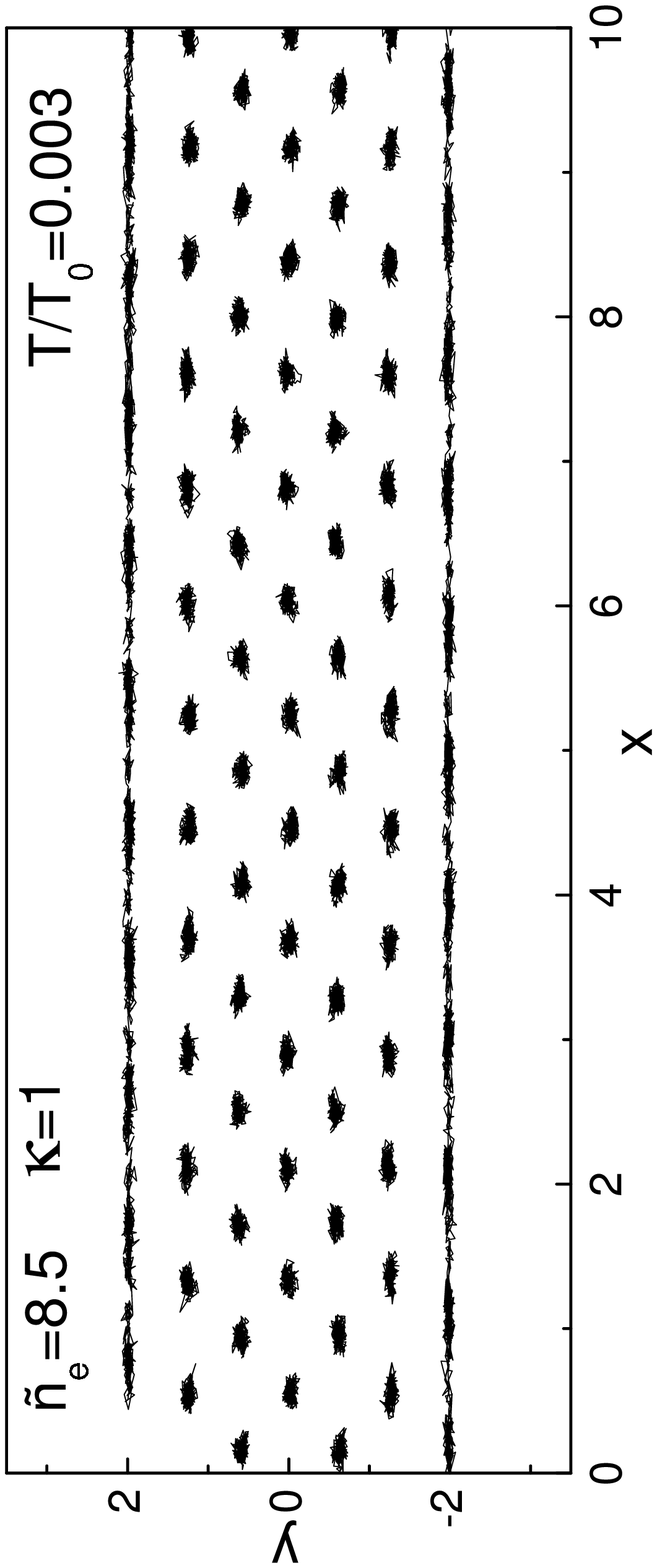}
\caption{Particle trajectories for $3 \times 10^7$ MC steps which
qualitatively illustrates the different melting behavior at the
boundaries due to the confining potential for $\kappa =1$.}
\label{figurename}
\end{center}
\end{figure}

\begin{figure}
\begin{center}
\includegraphics[width=10cm]{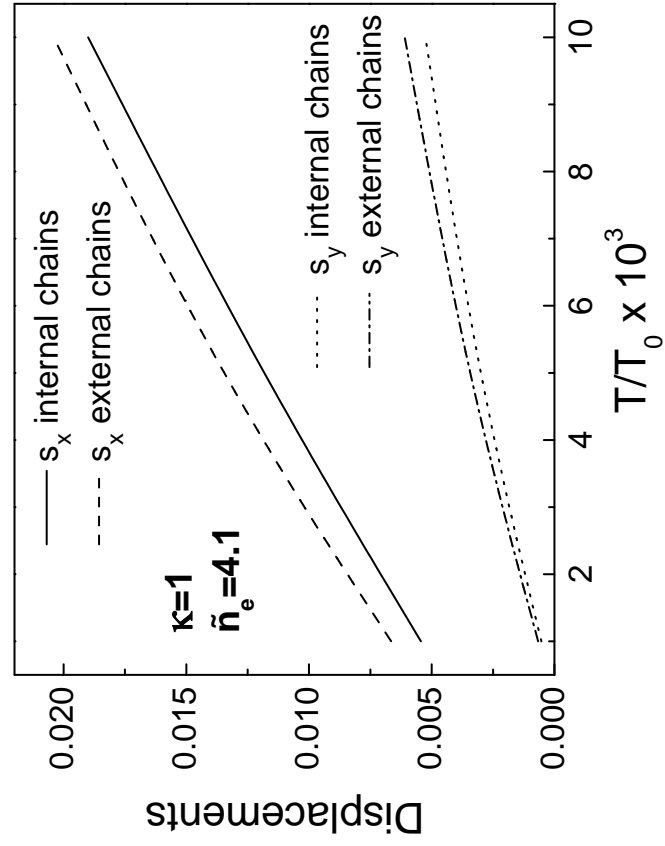}
\caption{Temperature dependence of the displacements in both the
unconfined and confined directions for external and internal
chains in the four-chain structure.} \label{figurename}
\end{center}
\end{figure}

\clearpage

\begin{figure}
\begin{center}
\includegraphics[width=10cm]{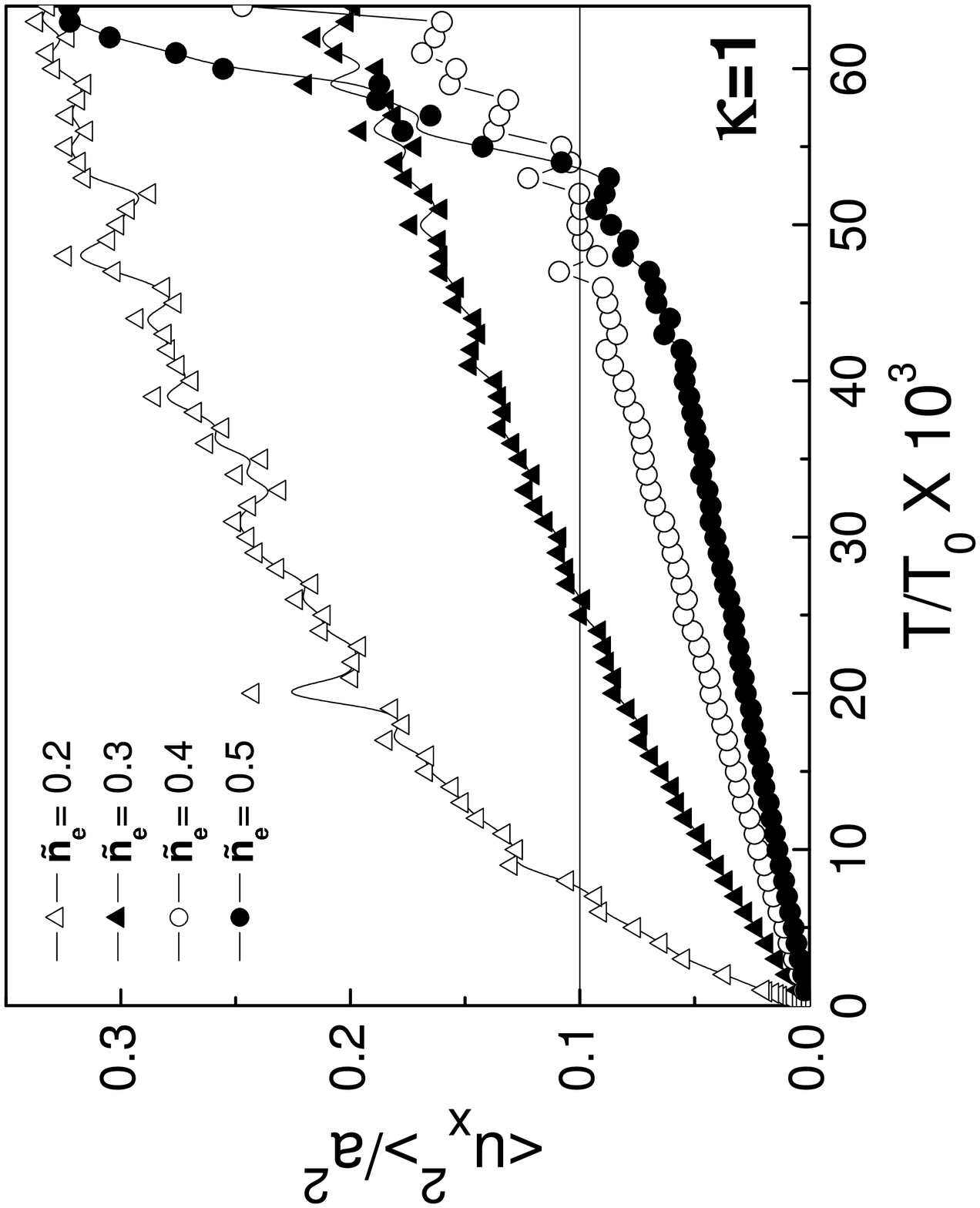}
\caption{The behavior of the Lindemann parameter, for the
single-chain regime at four different densities. It shows how the
linear regime at higher densities becomes sublinear at lower
densities.} \label{figurename}
\end{center}
\end{figure}

\clearpage

\begin{figure}
\begin{center}
\includegraphics[width=10cm]{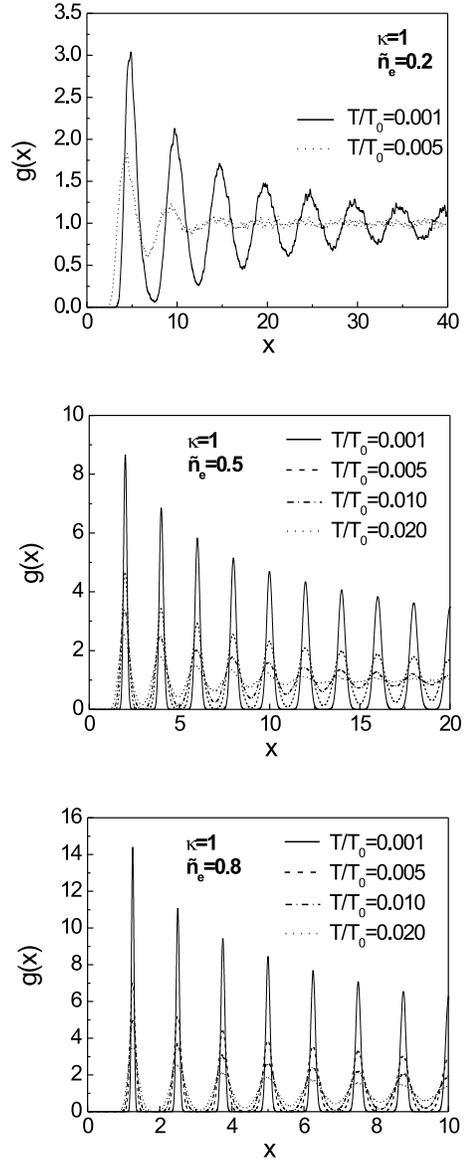}
\caption{The pair correlation function at different temperatures,
for three different densities, for the single chain
configuration.} \label{figurename}
\end{center}
\end{figure}

\clearpage

\begin{figure}
\begin{center}
\includegraphics[width=10cm]{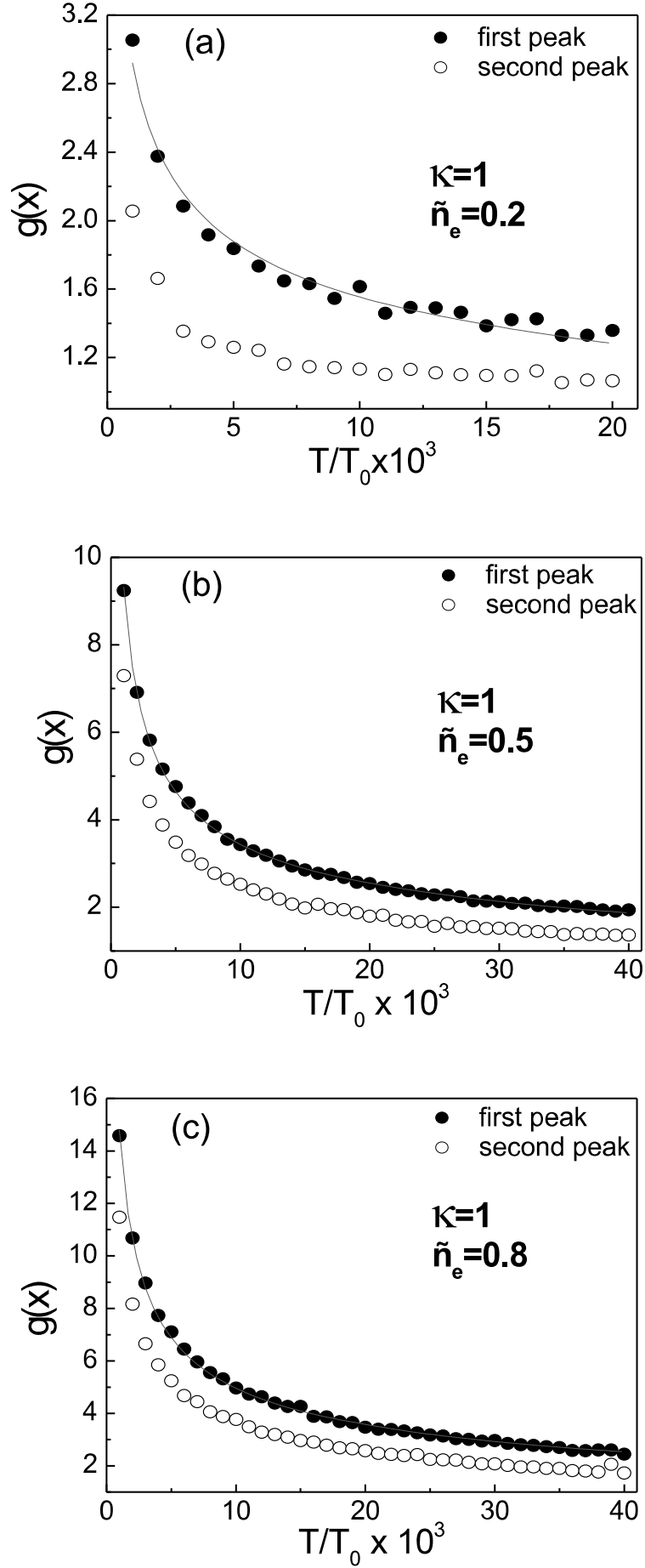}
\caption{The height of the first and second peaks of the pair
correlation function for the single chain as a function of
temperature for three different densities.The lines are the best
fits with the function $\alpha (T/T_0)^{-\beta}$.}
\label{figurename}
\end{center}
\end{figure}

\clearpage

\begin{figure}
\begin{center}
\includegraphics[width=10cm]{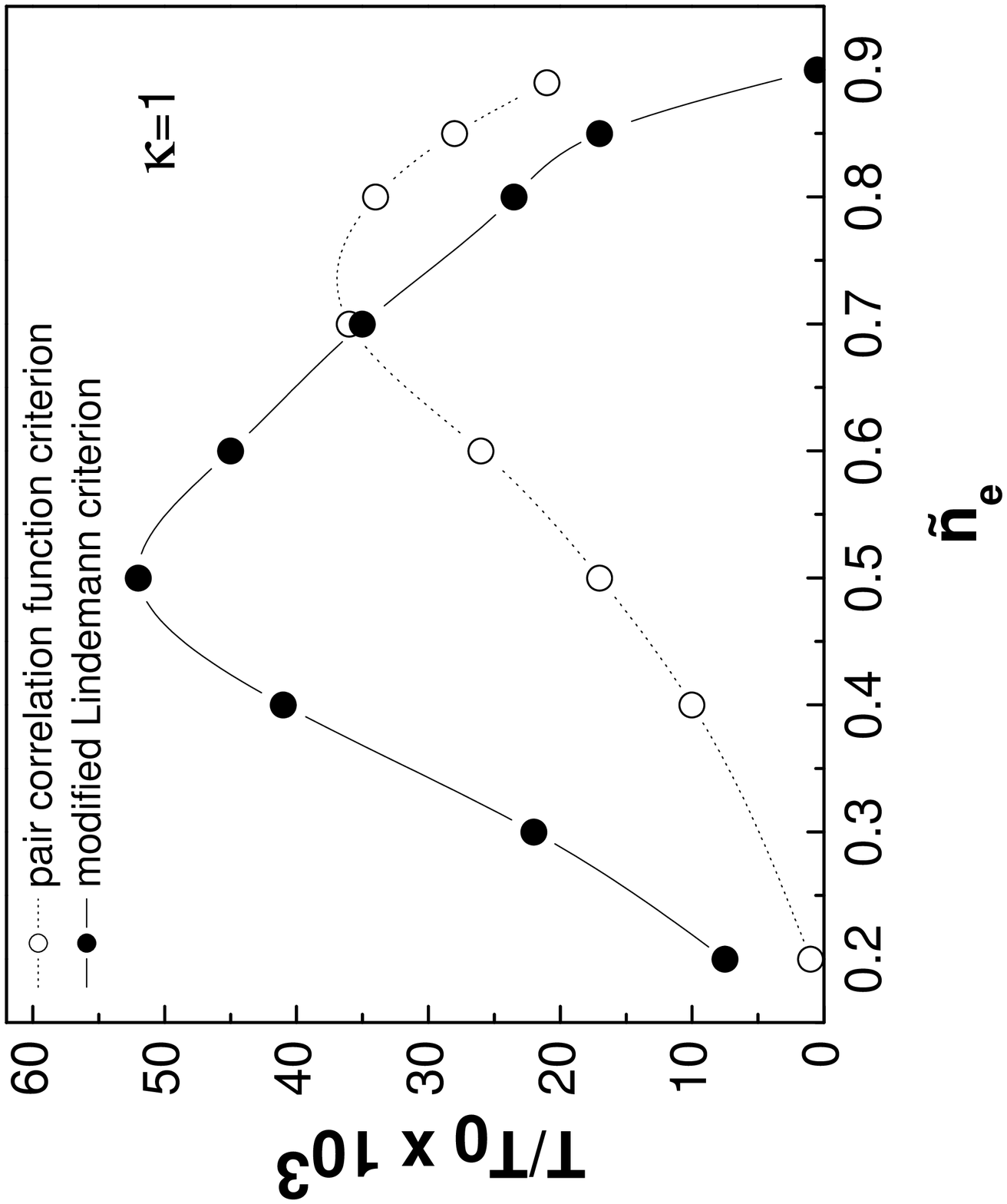}
\caption{The melting temperature for the single chain as obtained
from the two complementary criteria.} \label{figurename}
\end{center}
\end{figure}


\section*{References}

\begin{thebibliography} {*}

\bibitem[$\ast$]{giovanni} Email address: piacente@uia.ua.ac.be.

\bibitem[$\dagger$]{irina} Permanent address: Russian Academy of Science,
Institute of Theoretical and Applied Mechanics, Novosibirsk
630090, Russia. Email address: ischweig@itam.nsc.ru.

\bibitem[$\ddagger$]{joseph} Email address: betouras@uia.ua.ac.be.

\bibitem[$\circ$]{francois} Email address: francois.peeters@ua.ac.be.

\bibitem{qhestripes} M.~P. Lilly, K.~B. Cooper, J.~P. Eisenstein, L.~N. Pfeiffer,
 and K.~W. West, Phys. Rev. Lett. {\bf82}, 394
(1999).

\bibitem{hightc} E. Dagotto, T. Hotta, and A. Moreo, Physics Reports,
{\bf344}, 1 (2001); E.~W. Carlson, V.~J. Emery, S.~A. Kivelson, D.
Orgad, cond-mat/0206217.

\bibitem{glasson} P. Glasson, V. Dotsenko, P. Fozooni, M.~J. Lea, W. Bailey,
G. Papageorgiou, S.~E. Andresen, and A. Kristensen, Phys. Rev.
Lett. {\bf87}, 176802 (2001).

\bibitem{kovdrya} Yu. Z. Kovdrya, Low Temperature Physics {\bf29}, 77 (2003).

\bibitem{whitesides} G. M. Whitesides and A. D. Stroock, Physics Today {\bf54}, 42 (2001).

\bibitem{zahn} K. Zahn, R. Lenke, and G. Maret, Phys. Rev. Lett. {\bf82}, 2721 (1999).

\bibitem{chu} J.~H. Chu and L. I, Phys. Rev. Lett. {\bf72}, 4009 (1994).

\bibitem{wigner} E. Wigner, Phys. Rev. {\bf46}, 1002 (1934).

\bibitem{platzman} P.~M. Platzman and M.~I. Dykman, Science {\bf284}, 1967
(1999).

\bibitem{Mermin} N. D. Mermin, Phys. Rev. {\bf171}, 272 (1968).

\bibitem{Gann} R. C. Gann, S. Chakravarty, and G. V. Chester, Phys. Rev. B {\bf20},
326 (1979).

\bibitem{dubin} S.-J. Chen and D.~H.~E. Dubin, Phys. Rev. Lett. {\bf71},
2721 (1993).

\bibitem{schultz} H. J. Schultz, Phys. Rev. Lett. {\bf71}, 1864 (1993).

\bibitem{glazman} L.~I. Glazman, I.~M. Ruzin, and B.~I. Shklovskii,
Phys. Rev. B {\bf45}, 8454 (1992).

\bibitem{radzihovsky} L. Radzihovsky, E. Frey, and D. R. Nelson,
Phys. Rev. E {\bf63}, 031503 (2001) ; E. Frey, D.R. Nelson, and L.
Radzihovsky, Phys. Rev. Lett. {\bf83}, 2977 (1999).

\bibitem{magn1} E. Y. Andrei, G. Deville, D. C. Glattli,
F. I. B. Williams, E. Paris, and B. Etienne, Phys. Rev. Lett.
{\bf60}, 2765 (1988).

\bibitem{magn2} H. L. Stormer and R. L. Willett, Phys. Rev. Lett. {\bf62}, 972 (1989).

\bibitem{kthny} J.~M. Kosterlitz and D.~J. Thouless, J. Phys. C {\bf5}, L124
(1972); B.~I. Halperin and D.~R. Nelson, Phys. Rev. Lett. {\bf
41}, 121 (1978); A.~P. Young, Phys. Rev. B {\bf19}, 1855 (1979).

\bibitem{wei} Q. H. Wei, C. Bechinger, D. Rudhardt, and P. Leiderer,
Phys. Rev. Lett. {\bf81}, 2606 (1998).

\bibitem{bonsall} L. Bonsall and A.~A. Maradudin,Phys. Rev. B {\bf15}, 1959 (1977).

\bibitem{ziman} J.~M. Ziman, {\it Principles of the Theory of Solids}
(Cambridge University Press, Cambridge, 1972), p. 39-41.

\bibitem{fisher} D.~S. Fisher, Phys. Rev. B {\bf26}, 5009 (1982).

\bibitem{goldoni}G. Goldoni and F.~M. Peeters,Phys. Rev. B {\bf53}, 4591 (1996).

\bibitem{zigzag} L. Candido, J.~P. Rino, N. Studart, and F.~M. Peeters, J. Phys. : Cond. Matt. {\bf10}, 11627 (1998).

\bibitem{koulakov} A. A. Koulakov and B. I. Shklovskii, Phys. Rev. B {\bf57}, 2352 (1998).

\bibitem{landau} L.~D. Landau and E.~M. Lifshitz, {\it Elasticity Theory}
(Reed Educational and Professional Publishing, Exeter, 1996), Sec.
24.

\bibitem{vanlee} J. H. Van Leeuwen, J. Physique {\bf6}, 361 (1921).

\bibitem{sokolov} S.~S. Sokolov and O.~I. Kirichek, Low Temp. Phys. {\bf20}, 599 (1994).

\bibitem{chaplik} A. V. Chaplik, Sov. Phys. - JETP {\bf35}, 395(1972).

\bibitem{pines} D. Pine, {\it Elementary excitations in solids} (Addison-Weiley, N.Y., 1963), p. 13.

\bibitem{bedanov} V.~M. Bedanov and F.~M. Peeters, Phys. Rev. B {\bf49},
2667 (1994).

\bibitem{chklovskii} D.~B. Chklovskii, B.~I. Shklovskii, and L.~I. Glazman,
Phys. Rev. B {\bf46}, 4026 (1992).

\bibitem{andrei} E.~Y. Andrei, Ed., {\it Two dimensional Electron Systems on
Helium and Other Cryogenic Substrates} (Academic Press, New York,
1991).

\bibitem{mehrotra} K.~M.~S. Bajaj and R. Mehrotra, Physica B {\bf194-196},
1235 (1994).

\bibitem{cote} R. C$\hat{o}$t$\acute{e}$ and H.~A. Fertig, Phys. Rev. B {\bf48}, 10955 (1993).

\bibitem{vsch} I. V. Schweigert, V. A. Schweigert, and F. M. Peeters,
Phys. Rev. B {\bf 60}, 14665 (1999); Phys. Rev. Lett. {\bf82},
5293 (1999).

\bibitem{sokolov}  A. Valkering, J. Klier, and P. Leiderer, Physica B {\bf284}, 172 (2000).

\bibitem{kong} M. Kong, B. Partoens, and F.~M. Peeters, Phys. Rev. E {\bf67}, 021608 (2003).

\bibitem{nazarov} Y.~V. Nazarov, Europhys. Lett. {\bf32}, 443 (1995).

\end{thebibliography}
\end{document}